\documentclass[12pt]{iopart}
\usepackage[T1]{fontenc}
\usepackage[latin1]{inputenc}
\usepackage{color}
\usepackage{float}
\usepackage{bm}
\usepackage{amsbsy}
\usepackage{amstext}
\usepackage{amssymb}
\usepackage{graphicx}
\usepackage{esint}

\makeatletter
\usepackage{iopams}
\usepackage{setstack}

\usepackage{iopams}

\usepackage{cite}

\makeatother

\begin{document}

\title{Phase space methods for Majorana fermions}

\author{Ria~Rushin~Joseph$^{1}$, Laura~E~C~Rosales-Z\'arate$^{1,2}$
and Peter~D~Drummond$^{1}$ }

\address{$^{1}$ Centre for Quantum and Optical Science, Swinburne University
of Technology, Melbourne 3122, Australia}

\address{$^{2}$ Centro de Investigaciones en \'Optica A.C., Le\'on, Guanajuato
37150, M\'exico}

\ead{pdrummond@swin.edu.au}
\begin{abstract}
Fermionic phase space representations are a promising method for studying
correlated fermion systems. The fermionic Q-function and P-function
have been defined using Gaussian operators of fermion annihilation
and creation operators. The resulting phase-space of covariance matrices
belongs to the symmetry class D, one of the non-standard symmetry
classes. This was originally proposed to study mesoscopic normal-metal-superconducting
hybrid structures, which is the type of structure that has led to
recent experimental observations of Majorana fermions. Under a unitary
transformation, it is possible to express these Gaussian operators
using real anti-symmetric matrices and Majorana operators, which are
much simpler mathematical objects. We derive differential identities
involving Majorana fermion operators and an antisymmetric matrix which
are relevant to the derivation of the corresponding Fokker\textendash Planck
equations on symmetric space. These enable stochastic simulations
either in real or imaginary time. This formalism has direct relevance
to the study of fermionic systems in which there are Majorana type
excitations, and is an alternative to using expansions involving conventional
Fermi operators. The approach is illustrated by showing how a linear
coupled Hamiltonian as used to study topological excitations can be
transformed to Fokker-Planck and stochastic equation form, including
dissipation through particle losses. 
\end{abstract}


\submitto{\JPA }
\maketitle

\section{Introduction}

Due to the complexity arising from Pauli's exclusion principle, Fermi
statistics are a challenging topic in theoretical physics. As a result,
new approaches to understanding fermion physics are important. While
ideas from coherence theory play an important role in the physics
of bosons, they have not yet been widely used to study fermions. Probabilistic
methods are especially interesting, as they have the promise that
they can be efficiently sampled. Here we show, by using fermionic
Q-functions, that first order Fokker Plank equations can be obtained
both for linear coupling and for dissipation. This makes simulations
possible in regimes where topological excitations and Majorana fermions
are expected to occur. 

The control acquired in experiments of ultra-cold fermionic atomic
systems opens new areas that were not explored previously, including
Majorana fermions~\cite{Majorana:1937,Alicea:2012,Elliott:2015}
and the unitary limit of strong interactions~\cite{Bloch_RMP_2008_ManyBodyPhys_UG}.
These complement studies of strongly correlated fermions in condensed
matter physics, which are now investigating such exotic types of fermionic
excitation as well.

In particular, there has been a recent resurgence of interest in the
work of Majorana~\cite{Majorana:1937}. As well as the question of
fundamental Majorana fermions, which is still not completely resolved,
there is a growing interest in Majorana-like excitations and quasi-particles
in condensed matter physics \cite{fu2008superconducting}. There have
been several claims of experimental observations, typically in environments
involving coupling of semiconductors to superconductors~\cite{Mourik:2012,Nadj-Perge:2014,Xu:2015,Rokhinson:2012}.
Recent reviews regarding Majorana fermions focus on their possible
observations in nuclear, particle and solid states physics~\cite{Elliott:2015,Beenaker:2013}
or their possible relation to Random Matrix Theory~\cite{Beenakker:2015}.

Some of this work is also closely related to proposals for topological
insulators and quantum computing~\cite{DasSarma:2015}. While this
development is still speculative~\cite{Kitaev:2001,Stern:2013},
it does reveal an important issue. How can one evaluate topological
or related quantum technology proposals? Many studies of this type
omit crucial issues such as decoherence mechanisms. To understand
decoherence and fidelity, which are essential to quantum technology
applications, a method for treating such Hamiltonian terms theoretically
is needed. Neither mean field theory nor perturbation theory is sufficient
to treat these deeply quantum-mechanical systems. The use of exact
methods on an orthogonal basis lacks scalability; thus, other methods
will be needed. 

The general field of Majorana fermionic excitations in condensed matter
systems is growing~\cite{Alicea:2012}. The common problem that occurs
theoretically is that if one wishes to go beyond mean field theory,
then either one must work in a restricted Hilbert space of small dimension,
or else there are sign issues from using Monte-Carlo methods. Recent
methods in this direction, intended to overcome the sign problem,
include using the symmetry of the Hamiltonian properties of Lie groups
and Lie algebras~\cite{Wang:2015} or treating fermionic systems
interacting with spins~\cite{Huffman:2016}. A different sign free
approach is to consider the Majorana representation together with
Quantum Monte Carlo simulations~\cite{Li:2015,Li:2016,Wei:2016}.
Finally we point out that extremely powerful methods for classical
simulation of fermion chains with linear couplings are known to exist~\cite{Terhal:2002,Bravyi2012,Pedrocchi:2015},
originally based on work by Valiant~\cite{Valiant:2001}, although
this technique has not been used to treat particle loss or gain.

Just as the use of a coherent state basis and the corresponding phase-space
methods proved essential to understanding laser physics~\cite{WeidlichRiskenHaken1967},
it is useful to develop a similar coherent formalism for Majorana
physics. Phase space representations have been widely used in quantum
optics~\cite{Hillery_Review_1984_DistributionFunctions,Gardiner_Book_QNoise}
and ultracold atoms~\cite{Blakie:2008,Polkovnikov_2010_Phase_Rep_QDyn}
in order to study dynamics. The advantage of using phase-space methods
is that it is possible to obtain an exact Fokker Planck equation from
the time evolution equation of the density operator. The Fokker Planck
equation can be readily transformed to a stochastic differential equation~\cite{Carmichael:1999_Book,Gardiner_Book_SDE}.
An advantage of this method is that it is possible to sample the stochastic
equation probabilistically, allowing one to perform quantum dynamical
calculations for large number of modes and particles~\cite{drummond2016quantum}. 

These methods can be used to treat dissipation and losses. Such processes
occur in many physical systems and represent, for example, the loss
of atoms from a trap in the case of magnetically trapped atoms. For
electrons they may originate in the transfer of electrons from one
band to another. In general, such physical processes are part of the
quantum dynamics, and are described by the use of master equations.
Since these represent non-unitary processes on the subsystem of interest,
they require appropriate techniques able to represent the entire density
matrix. This is shown here through a simple example.

The purpose of these approaches is to enable the use of probabilistic
methods, which can overcome the exponential growth problem of large
many-body Hilbert spaces. An example is the positive Q-function. For
this representation, there have been applications using the SU(2)
coherent states~\cite{Arecchi_SUN,Gilmore:1975}, including simulations
of Bell violations~\cite{ReidqubitPhysRevA.90.012111}. Another application
of the dynamics of the Q-function was in the study of quantum correlations
of a degenerate optical parametric oscillator~\cite{Zambrini2003}. 

These methods have been recently extended to the fermionic raising
and lowering operators, resulting in fermionic analogs of the normally
ordered P-function and the anti-normally ordered Q-function for bosons.
Such phase space representations use as a basis the Gaussian operators~\cite{Corney_PD_JPA_2006_GR_fermions,Corney_PD_PRB_2006_GPSR_fermions,Corney_PD_PRL2004_GQMC_ferm_bos},
which are exponential of quadratic forms of creation and annihilation
operators. A related concept is the fermionic Gaussian state, for
which the density matrix is an exponential of quadratic forms of operators
defined by a correlation matrix. Some applications of the fermionic
Gaussian operators include the study of the Fermi Hubbard model~\cite{Imada_2007_GBMC,Corboz_Chapter_PhaseSpaceMethodsFermions,Corney_PD_PRL2004_GQMC_ferm_bos},
and fermionic molecular dissociation~\cite{Ogren:2010_Qdynamics_fermions_MolDiss,Ogren2011_fermiondynamics}. 

It was proved recently that the fermionic Q-function exists as a complete
probabilistic representation~\cite{FermiQ}. Hence, it is important
to develop this method further.  Earlier theoretical work used a Gaussian
basis of Fermi operators~\cite{Corney_PD_JPA_2006_GR_fermions,Corney_PD_PRB_2006_GPSR_fermions,Corney_PD_PRL2004_GQMC_ferm_bos}.
By transforming the basis in terms of Majorana operators, the resulting
differential identities can help to treat Majorana physics also. 

The fermionic Gaussian operators that we consider belong to the non-standard
symmetry class D defined by Altland and Zirnbauer~\cite{Altland_Zirnbauer:1997}.
This non-standard symmetry class was originally proposed to study
mesoscopic normal-metal-superconducting hybrid structures, which is
exactly the type of structure that has led to recent experimental
observations of Majorana quasi-particles. We show that, under a unitary
transformation, we can write the class D operators in terms of an
anti-symmetric matrix and Majorana variables~\cite{Majorana:1937},
and obtain a positive phase-space representation. Since the distribution
is now defined in terms of the Majorana operators, we expect it will
be useful in investigating the recently discovered Majorana fermionic
quasi-particles in hybrid superconductor  or ultra-cold atomic systems~\cite{Mourik:2012,Nadj-Perge:2014,Liu:2012}.
Other symmetries can also be used to identify classes of operator
transformations, and to reduce the resulting dimensionality. These
are not treated here, but allow this work to be extended.

There has been much work exploring the connection of Class D symmetry
with topological and Majorana-like excitations~\cite{Beenaker:2013,Lutchyn:2010,Bagrets:2012,DeGottardi:2013,Gibertini:2013,Morimoto:2013,Budich:2013,Chiu:2016,Hegde:2016}.
This relationship therefore has a strong physical underpinning. Moreover,
the use of conventional methods such as mean field theory is not applicable
when there are strong quantum correlations. Therefore, our approach
is to investigate other mathematical relations involving differential
identities, that can help these investigations in future. These identities
allow the symmetry class to be utilized in the operator representation.
The end result is that one can treat exponentially complex Hilbert
spaces as dynamical equations in a continuous phase-space, which are
generally simpler to solve.

In this communication, we derive differential identities for the Gaussian
phase space representation using Majorana operators. These identities
are relevant to the general use of a Gaussian basis and to the fermionic
Q and P-function. Differential identities allow simulation of the
time evolution either in real or imaginary time. The latter approach
has already been used to investigate the ground state of the Fermi
Hubbard model \cite{Imada:2007_2DHM_Superconductivity,Imada_2007_GBMC}.
Thus, these differential identities are potentially relevant to the
dynamics of Majorana systems or to systems that have Majorana operator
excitations. We expect that using methods based on Majorana operators
will further help to study other Fermi systems as well, because the
class D symmetry has a fundamentally simpler form using these operators.

This paper is organized as follows: In section~\ref{sec:Gaussian-Operators}
we review the Gaussian phase space representation as well as we introduced
the Majorana Gaussian operators. Section~\ref{sec:Majorana-differential-properties}
presents our main result, which is the differential identities for
the Majorana operators. These are given for the un-normalized and
normalized form of the Majorana Gaussian operator. We obtain identities
for both ordered and unordered Majorana operator products. In section~\ref{sec:Phase-space-representations}
we present the corresponding phase space representations with Fermi
and Majorana operators. We present the time evolution equation for
the Majorana Q-function for one of these differential equations in
section~\ref{sec:Time-Evolution-MQf}. In section~\ref{sec:TimeEvOpenSyst},
we discuss the time evolution of the Q-function for an open quantum
system. Finally, a summary of our results and conclusions are given
in section~\ref{sec:Summary_Conclusion}. 

\section{Fermionic Gaussian operators\label{sec:Gaussian-Operators}}

Fermionic phase space representations that use a phase-space of Grassmann
variables are known~\cite{Cahill_Glauber_fermions_1999}. While this
is useful for analytic calculations, it is an exponentially complex
problem to directly represent the Grassmann variables computationally.
Therefore, in order to avoid such exponential complexity, it is helpful
to map the fermion problem into a phase-space of covariances, which
are more computationally tractable. One method to achieve this is
to use the fermionic Gaussian operators as a basis~\cite{Corney_PD_JPA_2006_GR_fermions,Corney_PD_PRB_2006_GPSR_fermions,ResUnityFGO:2013,FermiQ}.
In this case, the phase space is a space of covariance matrices, which
are complex variables. We emphasise that through using identities,
any correlation order can be calculated, including the extremely high
orders that can occur in some quantum technology and mesoscopic physics
applications~\cite{Opanchuk:2014,ReidqubitPhysRevA.90.012111}. An
alternative method is to derive the covariance matrix equations of
motion from Grassmann variable equations \cite{dalton2016grassmann},
which is not treated here.

In this section, we review the properties of the fermionic Gaussian
operators in the notation used in the paper. These were first introduced
as variational BCS states~\cite{bogolyubov1958zh,valatin1958comments},
and their algebraic properties as operator transformations have been
studied extensively~\cite{Balian_Brezin_Transformations,Altland_Zirnbauer:1997}.
We consider a fermionic system that can be decomposed into $M$ modes,
where $\hat{\bm{a}}$ is defined as a vector of $M$ annihilation
operators, while $\hat{\bm{a}}^{\dagger}$ is the corresponding vector
of $M$ creation operators. The Fermi operators, $\hat{a}_{i}$ and
$\hat{a}_{j}^{\dagger}$, obey the usual anti commutation relations:
\begin{equation}
\left\{ \hat{a}_{i},\hat{a}_{j}^{\dagger}\right\} =\delta_{ij},\qquad\left\{ \hat{a}_{i},\hat{a}_{j}\right\} =0.\label{eq:FerComm}
\end{equation}
 The single mode number operator is $\hat{n}=\widehat{a}^{\dagger}\widehat{a}$,
and the action of the operators on the single-mode number states $\mid n\rangle$
is that ${\displaystyle \widehat{a}^{\dagger}\mid n\rangle=}\left(1-n\right)|n+1\rangle$,
and ${\displaystyle \widehat{a}\mid n\rangle=}n|n-1\rangle$.

Using the Fermi operators, we define a $2M$ extended vector of creation
and annihilation operators $\hat{\underline{a}}=\left(\hat{\bm{a}}^{T},\hat{\bm{a}}^{\dagger}\right)^{T},$
with the corresponding adjoint vector defined as $\hat{\underline{a}}^{\dagger}=\left(\hat{\bm{a}}^{\dagger},\hat{\bm{a}}^{T}\right)=\left(\hat{a}_{1}^{\dagger},\ldots,\hat{a}_{M}^{\dagger},\hat{a}_{1},\ldots,\hat{a}_{M}\right)$.
We use the following notation throughout the paper: $M$ vectors are
in bold type as $\hat{\bm{a}}$, $2M$ vectors are denoted with a
single underline as $\hat{\underline{a}}$, $M\times M$ matrices
are in bold type as $\mathbf{I}$, and $2M\times2M$ matrices are
denoted with a double underline as $\underline{\underline{I}}$. 

The un-normalized fermionic Gaussian operators $\hat{\Lambda}_{f}^{u}$
are defined here~\cite{Corney_PD_JPA_2006_GR_fermions,Corney_PD_PRB_2006_GPSR_fermions}
as the set of all possible normally-ordered Gaussian functions of
these $2M$ dimensional operator vectors,

\begin{equation}
\hat{\Lambda}_{f}^{u}\Biggl(\underline{\underline{\mu}}\Biggr)=:\exp\Biggl[-\hat{\underline{a}}^{\dagger}\underline{\underline{\mu}}\hat{\underline{a}}/2\Biggr]:.
\end{equation}
Similar definitions were subsequently introduced elsewhere~\cite{Kraus:2009,Eisler:2015},
except without normal ordering. Throughout the paper, normal ordering
applies to the annihilation and creation operators, and is denoted
by $:\ldots:$. This includes a sign change for every fermion operator
swap, hence $:\hat{a}_{i}\hat{a}_{j}^{\dagger}:=-\hat{a}_{j}^{\dagger}\hat{a}_{i}$.
Anti-normal ordering is denoted by $\left\{ \ldots\right\} $, hence
$\left\{ \hat{a}_{j}^{\dagger}\hat{a}_{i}\right\} =-\hat{a}_{i}\hat{a}_{j}^{\dagger}$.
We will also use nested orderings~\cite{Corney_PD_JPA_2006_GR_fermions}.
In this case the outer ordering does not reorder the inner one, but
the sign changes still take place for each swap - for example $\left\{ :\hat{O}\hat{b}_{i}^{\dagger}:\hat{b}_{j}\right\} =-\hat{b}_{j}\hat{b}_{i}^{\dagger}\hat{O}$.
A convenient notation is to introduce a basis of  unit trace fermionic
operators $\hat{\Lambda}_{f}$, where:

\begin{equation}
\hat{\Lambda}_{f}=\sqrt{\det\left[\rmi\underline{\underline{\sigma}}\right]}:\exp\left[-\hat{\underline{a}}^{\dagger}\left(\underline{\underline{\sigma}}^{-1}-2\underline{\underline{J}}\right)\hat{\underline{a}}/2\right]:\,.\label{eq:NormalizedGO_sigma}
\end{equation}
 Here $\underline{\underline{J}}$ is a matrix square root of the
identity:
\begin{equation}
\underline{\underline{J}}=\left[\begin{array}{cc}
-\mathbf{I} & \mathbf{0}\\
\mathbf{0} & \mathbf{I}
\end{array}\right].\label{eq:I_GO}
\end{equation}

The importance of expressing the Gaussian operators in this form is
that these Gaussian operators using annihilation and creation operators
are the basis for the fermionic Q and P-functions. The matrices $\mathbf{0}$
and $\mathbf{I}$ are the $M\times M$ zero and identity matrices,
respectively, and $\underline{\underline{\sigma}}=\left(\underline{\underline{\mu}}+2\underline{\underline{J}}\right)^{-1}$
is the \emph{covariance} matrix, which can be expressed in terms of
an $M\times M$ matrix $\mathbf{n}$ and a complex antisymmetric $M\times M$
matrix $\mathbf{m}$:
\begin{equation}
\underline{\underline{\sigma}}=\left(\underline{\underline{\mu}}+2\underline{\underline{J}}\right)^{-1}=\left[\begin{array}{cc}
\mathbf{n}^{T}-\mathbf{I} & \mathbf{m}\\
\mathbf{m}^{+} & \mathbf{I}-\mathbf{n}
\end{array}\right].\label{eq:Sigma-decomposition}
\end{equation}

We use the notation $\mathbf{n}^{T}$ to indicate a transpose, $\mathbf{m}^{*}$
a complex conjugate, and $\mathbf{m}^{\dagger}$ a conjugate transpose.
The most general fermionic Gaussian operator is non-hermitian. In
this case $\mathbf{n}$ is not hermitian, and $\mathbf{m}^{+}$, $\mathbf{m}$
are independent antisymmetric complex matrices. To obtain hermitian
Gaussian operators, one must impose the additional restrictions that
$\mathbf{m}^{+}=\mathbf{m}^{\dagger}$ and $\mathbf{n}=\mathbf{n}^{\dagger}$~\cite{Corney_PD_PRB_2006_GPSR_fermions}.
The $\bm{n}$ and $\bm{m}$ notation is used because these terms in
the covariance matrix of a hermitian Gaussian density matrix are equal
to the normal and anomalous Green's functions, which are ubiquitous
in fermionic many-body theory.  We note that the covariance matrix
correspond to the first-order correlations of the Gaussian operators
in normally ordered form, since $\Tr\left[:\hat{\underline{a}}\hat{\underline{a}}^{\dagger}\hat{\Lambda}_{f}\right]=\underline{\underline{\sigma}}$~\cite{Corney_PD_JPA_2006_GR_fermions}.

The Gaussian operators have a more widespread utility than just their
obvious value in treating Gaussian states, which are the ground state
solutions of quadratic Hamiltonians. The general fermionic Gaussian
operators are a complete basis for all fermionic density matrices,
including those that are not Gaussian, through taking linear combinations
of Gaussian operators. On the other hand, the hermitian fermionic
Gaussian operators can be regarded as physical density matrices by
themselves.

Several nonstandard symmetry classes were defined by Altland and Zirnbauer~\cite{Altland_Zirnbauer:1997}.
Here we focus on Class D, which corresponds to cases which do not
have time-reversal or spin-rotation symmetry. The general hermitian
Gaussian operators have the nonstandard Class D symmetry~\cite{FermiQ}. 

\subsection{Majorana Gaussian operators \label{subsec:Majorana-Gaussian-operators}}

The work of Majorana~\cite{Majorana:1937} showed that a more symmetric
form of quantization was obtainable if the Fermi operators are transformed
to an hermitian Majorana fermion operator basis. We use the normalization
that is most common in modern papers, where the Majorana operators
are obtained using a matrix transformation of the usual Fermi annihilation
and creation operators~\cite{Elliott:2015},

\[
\hat{\underline{\gamma}}=\left[\begin{array}{c}
\hat{\bgamma}_{(1)}\\
\hat{\bgamma}_{(2)}
\end{array}\right]=\underline{\underline{U_{0}}}\hat{\underline{a}}\,\,.
\]
 This uses a matrix $\underline{\underline{U_{0}}}$, given by~\cite{Balian_Brezin_Transformations}:
\begin{equation}
\underline{\underline{U_{0}}}=\left[\begin{array}{cc}
\mathbf{I} & \mathbf{I}\\
-\rmi\mathbf{I} & \rmi\mathbf{I}
\end{array}\right],\label{eq:U0}
\end{equation}
which is not unitary, since $\underline{\underline{U_{0}}}^{\dagger}=2\underline{\underline{U_{0}}}^{-1}$,
although $\underline{\underline{U_{0}}}/\sqrt{2}$ is unitary. This
definition corresponds to introducing hermitian quadrature operators,
$\hat{\bgamma}_{(i)}$, as:

\begin{eqnarray}
\hat{\bgamma}_{(1)} & = & \left(\hat{\bm{a}}+\hat{\bm{a}}^{\dagger}\right)\nonumber \\
\hat{\bgamma}_{(2)} & = & -\rmi\left(\hat{\bm{a}}-\hat{\bm{a}}^{\dagger}\right),\label{eq:p-var}
\end{eqnarray}
which implies that the $\hat{\gamma}$ variables have the following
anti-commutation relation~\cite{Elliott:2015,Beenaker:2013,Beenakker:2015}.:
\begin{equation}
\left\{ \hat{\gamma}_{i},\hat{\gamma}_{j}\right\} =2\delta_{ij}.\label{eq:anti-com_MajoranaOp}
\end{equation}
The relationship can be inverted, and the extended ladder operators
can be expressed as:
\begin{equation}
\widehat{\underline{a}}=\underline{\underline{U_{0}}}^{-1}\hat{\underline{\gamma}}.\label{eq:OpMajUn}
\end{equation}
 Using the transformation $\underline{\underline{U_{0}}}$ it is possible
to write an un-normalized Gaussian operator as:
\begin{equation}
\hat{\Lambda}^{u}\Biggl(\underline{\underline{Y}}\Biggr)=:\exp\Biggl[\frac{\rmi}{2}\hat{\underline{\gamma}}^{T}\underline{\underline{Y}}\hat{\underline{\gamma}}\Biggr]:.\label{eq:GOY}
\end{equation}
We will call the above operators the un-normalized Majorana Gaussian
operators. Here we have defined:
\begin{equation}
\underline{\underline{Y}}=\frac{\rmi}{2}\underline{\underline{U_{0}}}\underline{\underline{\mu}}\underline{\underline{U_{0}}}^{-1}.\label{eq:YMat}
\end{equation}
Provided that $\underline{\underline{\mu}}$ has the class D symmetry
properties identified above, the matrix $\underline{\underline{Y}}$
is a real antisymmetric matrix. In the single mode case, one can define
$Y\equiv Y_{12}$, and since all normally ordered powers beyond first
order will vanish, 

\begin{equation}
\hat{\Lambda}^{u}\Biggl(Y\Biggr)=1+iY:\hat{\gamma}_{1}\hat{\gamma}_{2}:=1+2\hat{n}Y\,.
\end{equation}

\subsection{Unit trace Gaussian operators}

We would like to express the unit trace Gaussian operators in terms
of Majorana operators, together with a matrix $\underline{\underline{X}}$
that plays the role of a Majorana covariance. Covariance matrices
and Majorana variables have been used in order to study topological
edge states correlations in free and interacting fermion systems~\cite{Meichanetzidis:2016}.
They have also been used to study pairing in fermionic systems~\cite{Kraus:2009}. 

For this purpose, we define the following real antisymmetric matrices:
\begin{equation}
\underline{\underline{X}}=\rmi\underline{\underline{U_{0}}}\left[\underline{\underline{J}}-2\underline{\underline{\sigma}}\right]\underline{\underline{U_{0}}}^{-1},\label{eq:XMat}
\end{equation}
and an antisymmetric real 'identity-like' matrix 
\begin{equation}
\underline{\underline{\mathcal{I}}}=\rmi\underline{\underline{U_{0}}}\underline{\underline{J}}\underline{\underline{U_{0}}}^{-1},\label{eq:UnitFI}
\end{equation}
which can be written explicitly as;

\[
\underline{\underline{{\cal I}}}=\left[\begin{array}{cc}
\mathbf{0} & \mathbf{\mathbf{I}}\\
-\mathbf{I} & \mathbf{0}
\end{array}\right].
\]
Equation~(\ref{eq:Sigma-decomposition}) gives a relationship between
the variance matrix $\underline{\underline{\sigma}}$ and the $\underline{\underline{\mu}}$
matrix. In terms of the corresponding antisymmetric matrices $\underline{\underline{X}}$
and $\underline{\underline{Y}}$, this relationship is given by:
\begin{equation}
\underline{\underline{X}}=\underline{\underline{\mathcal{I}}}+\left(\underline{\underline{Y}}+\underline{\underline{\mathcal{I}}}\right)^{-1}.\label{eq:RelXY}
\end{equation}
In the single-mode case, one has that:
\begin{equation}
\underline{\underline{X}}=\left[\begin{array}{cc}
0 & 1-1/\left(1+Y\right)\\
1/\left(1+Y\right)-1 & 0
\end{array}\right].
\end{equation}
In order to define the normalized Gaussian operators using Majorana
operators, we first consider the normalization factor. Since the determinant
is invariant under unitary transformations we can write it as follows:
\begin{equation}
\sqrt{\det\left[\rmi\underline{\underline{\sigma}}\right]}=\frac{1}{2^{M}}\sqrt{\det\left[\underline{\underline{\mathcal{I}}}-\underline{\underline{X}}\right]}
\end{equation}
Therefore the unit-trace Majorana Gaussian operator in a variance
form is:
\begin{equation}
\hat{\Lambda}\left(\underline{\underline{X}}\right)=N\left(\underline{\underline{X}}\right):\exp\Biggl[\textcolor{blue}{-}\rmi\hat{\underline{\gamma}}^{T}\left[\underline{\underline{\mathcal{I}}}-\left(\underline{\underline{X}}-\underline{\underline{\mathcal{I}}}\right)^{-1}\right]\hat{\underline{\gamma}}/2\Biggr]:,\label{eq:MajOpX}
\end{equation}
where 
\begin{equation}
N\left(\underline{\underline{X}}\right)=\frac{1}{2^{M}}\sqrt{\det\left[\underline{\underline{\mathcal{I}}}-\underline{\underline{X}}\right]}.\label{eq:NormGOX}
\end{equation}
In the single-mode case, with $X=X_{12},$ the normalization factor
is:

\begin{equation}
N\left(X\right)=\frac{1}{2(1+Y)}=\frac{1-X}{2},
\end{equation}
where the domain boundaries are such that $-1<X<1$, and: 
\begin{equation}
\hat{\Lambda}\left(X\right)=\frac{1-X}{2}+\hat{n}X\,.
\end{equation}
We note that there is a particle-hole symmetry, since $\hat{\Lambda}\left(X\right)$
is invariant under the transformation $\hat{a}\rightarrow\hat{a}^{\dagger}$,
$\hat{a}^{\dagger}\rightarrow\hat{a}$, $X\rightarrow-X$.

\subsection{Classical domains }

After transformation to Majorana form, there are two possible fermionic
representations with different phase-spaces. The fermionic P-representation
corresponds to the complex antisymmetric matrices in Majorana space,
while the Q-representation corresponds to the real antisymmetric matrices.

As we show explicitly below, the natural phase-space of the Majorana
Q-function is bounded. This is because the Gaussian basis that is
used in the construction of the Q-function depends on an antisymmetric
matrix $\underline{\underline{X}}$, which exists in one of the irreducible
bounded symmetric domains of Cartan~\cite{Cartan:1926,Cartan:1927Bull,Cartan:1935}.
In the complex case, these symmetric spaces are called the classical
domains by Hua~\cite{Hua_Book_harmonic_analysis,Xu:2015}. The Gaussian
basis has deep links to Lie group theory, since we can define a $2M\times2M$
matrix $\underline{\underline{T}}=\exp\left[\underline{\underline{X}}\right]$
such that $\underline{\underline{T}}^{T}\underline{\underline{T}}=\underline{\underline{I}}$,
where $\underline{\underline{I}}$ is the usual identity matrix. The
$\underline{\underline{T}}$ matrices belong to the orthogonal group
$O(2M)$, which is a Lie group. In the case that $\det\underline{\underline{T}}=1,$
the group is $SO(2M)$. The corresponding Lie algebra of this Lie
group is isomorphic to the real $2M\times2M$ antisymmetric matrices. 

The classical domains ~\cite{Cartan:1935} are given in terms of
complex matrices $\underline{\underline{Z}}$. This leads to a consideration
of the four classical domains, which in the classification of Hua~\cite{Hua_Book_harmonic_analysis}
are defined as~\cite{Helgason_book_LieGroup,Caselle_Magnea:2004_ReviewRMT_SymSpaces,Cartan:1927Bull,XuComplexDom}:
\begin{enumerate}
\item The domain $\mathcal{R}_{I}$ of $m\times n$ complex matrices with:
$\underline{\underline{I}}^{(m)}-\underline{\underline{Z}}\underline{\underline{Z}}^{\dagger}>0.$
\item The domain $\mathcal{R}_{II}$ of $n\times n$ symmetric complex matrices
with: $\underline{\underline{I}}-\underline{\underline{Z}}\underline{\underline{Z}}^{*}>0.$
\item The domain $\mathcal{R}_{III}$ of $n\times n$ skew-symmetric (anti-symmetric)
complex matrices with: $\underline{\underline{I}}+\underline{\underline{Z}}\underline{\underline{Z}}^{*}>0.$
\item The domain $\mathcal{R}_{IV}$ of $n$-dimensional vectors $z=\left(z_{1,}z_{2},\ldots,z_{n}\right),$
where $z_{k}$ are complex numbers, satisfying: $\left|zz^{T}\right|+1-2zz^{T}>0,\qquad\left|zz^{T}\right|<1.$
\end{enumerate}
The most general Majorana Gaussian operators correspond to the complex
domain $\mathcal{R}_{III}$, for $n=2M$, which we refer to as $\mathcal{D}_{C}.$
The subspace that corresponds to the hermitian Hamiltonians that we
use here has a real phase space of $M(2M-1)$ dimensions, or $\mathcal{D}_{R}$,
which is the real subspace of $\mathcal{R}_{III}$ . 

In our notation, the classical domain boundary of the space of the
most general complex antisymmetric matrices $\underline{\underline{Z}}$
considered here is given by:
\begin{equation}
\underline{\underline{R}}=\underline{\underline{I}}+\underline{\underline{Z}}\underline{\underline{Z}}^{*}>0.\label{eq:BoundXM}
\end{equation}
Since $\underline{\underline{R}}$ is positive-definite, and $\underline{\underline{Z}}\underline{\underline{Z}}^{*}=-\underline{\underline{Z}}\underline{\underline{Z}}^{\dagger}$
is negative definite, we deduce that each element of $\underline{\underline{Z}}$
is bounded, which implies that $\sum_{j}\left|Z_{ij}\right|^{2}<1$
and $\left|Z_{ij}\right|<1$. The classical domain boundary of the
real antisymmetric matrices $\underline{\underline{X}}$ is the same,
except for the additional requirement that the elements are real. 

This restriction of the domain $\mathcal{R}_{III}$ corresponds to
the hermitian Cartan symmetric space DIII~\cite{Cartan:1926,Cartan:1927Bull}.
Altland and Zirnbauer~\cite{Altland_Zirnbauer:1997} classified non-standard
symmetry classes for fermionic systems. Here the classification is
made by considering the Lie algebra which the symmetry class belongs
to \cite{Altland_Zirnbauer:1997,HeinznerZirnbauer:2005_SymmetryClasses_disorder}.
In this notation, their symmetry class $D$ corresponds to $D_{N}\equiv so(2N)$~\cite{Altland_Zirnbauer:1997,Caselle_Magnea:2004_ReviewRMT_SymSpaces,HeinznerZirnbauer:2005_SymmetryClasses_disorder}.
The Lie algebra is isomorphic to the real antisymmetric $2N\times2N$
matrices. 

The boundary condition can also be written in the real case as: 
\begin{equation}
\underline{\underline{X}}^{+}\underline{\underline{X}}^{-}>0,
\end{equation}
where we define for convenience, 
\begin{equation}
\underline{\underline{X}}^{\pm}\equiv\underline{\underline{X}}\pm\underline{\underline{\mathcal{I}}}.\label{eq:Xpm}
\end{equation}

For the unordered operator identities derived below, it is simpler
to transform the resulting phase-space variable to a modified form
such that:
\begin{equation}
\underline{\underline{x}}=\underline{\underline{{\cal I}}}\underline{\underline{X}}^{T}\underline{\underline{{\cal I}}}.\label{eq:x}
\end{equation}

We also introduce $\underline{\underline{x}}^{\pm}=\underline{\underline{x}}\pm\rmi\underline{\underline{I}}$,
and one can prove that in this case the classical domain bounds are
unchanged, with boundaries at: 
\begin{equation}
\underline{\underline{I}}+\underline{\underline{x}}^{2}=\underline{\underline{x}}^{+}\underline{\underline{x}}^{-}>0\,.
\end{equation}

\subsection{Resolution of identity}

The definition of a Q-function first requires one to have an expression
for the resolution of the Hilbert space identity operator. In order
to obtain the most general form of the resolution of identity in the
fermion case, we define $\hat{\Lambda}^{N}\left(\underline{\underline{X}}\right)$
as the following Majorana Gaussian basis:
\begin{eqnarray}
\hat{\Lambda}^{N}\left(\underline{\underline{X}}\right) & = & \frac{1}{{\cal N}}\hat{\Lambda}\left(\underline{\underline{X}}\right)S\left(\underline{\underline{X}}^{2}\right),\label{eq:MajGBasis}
\end{eqnarray}
with $\hat{\Lambda}\left(\underline{\underline{X}}\right)$ given
in (\ref{eq:MajOpX}),  $S\left(\underline{\underline{X}}^{2}\right)$
is an even, positive scaling function, and $\mathcal{N}$ is a normalization
constant that ensures that the identity expansion is normalized. We
consider the following form for the scaling factor $S$, which vanishes
at the phase-space boundaries:
\begin{equation}
S=\det\left[\underline{\underline{I}}+\underline{\underline{X}}^{2}\right]^{k/2}.
\end{equation}
Following the methods developed earlier \cite{ResUnityFGO:2013} for
the complex variable Q-function, one can then prove that:
\begin{equation}
\hat{I}=\int_{\mathcal{D}_{R}}\hat{\Lambda}^{N}\left(\underline{\underline{X}}\right)\rmd X\,,
\end{equation}
where the integration is over the bounded real classical domain of
antisymmetric matrices, and $dX=\prod_{1\text{\ensuremath{\le}}j<k\text{\ensuremath{\le}}2M}dX_{jk}.$
The value of the normalization constant is obtained by noting that:
\begin{equation}
{\cal N}=2^{-M}\int_{\mathcal{D}_{R}}S\left(\underline{\underline{X}^{2}}\right)\rmd X\,.
\end{equation}
In the limit that $k\rightarrow0$, the constant $\mathcal{N}$ is
related to the volume of the real classical Hua type $III$ domain
\cite{Hua_Book_harmonic_analysis,Mehta_RM_book,Forrester_book:2010}
using matrix polar coordinate methods, and is given by:
\begin{eqnarray}
\mathcal{N} & = & \frac{\pi^{M\left(M-1/2\right)}}{2^{M}}\prod_{j=1}^{M}\frac{\Gamma\left(k+j\right)}{\Gamma\left(k+M+j-\frac{1}{2}\right)}.\label{eq:normalization factor-1}
\end{eqnarray}
Thus, for example, in the case of $M=1$, one obtains $\mathcal{N}=1$.
We note that these results hold equally well when using the transformed
variable $\underline{\underline{x}}$, which has an identical measure,
determinant and integration domain. In the single mode case, the two
forms are exactly the same.

\section{Majorana differential identities\label{sec:Majorana-differential-properties}}

The main result of the paper is the derivation of differential identities
using Majorana operators and antisymmetric matrices. It is useful
to derive differential identities for the normalized Majorana Gaussian
operators in terms of the matrices $\underline{\underline{X}}$, because
they are related to the covariance matrices $\underline{\underline{\sigma}}$.
The fermionic Q-function uses as a basis the Gaussian operators expressed
in a symmetric form of these variables. Calculations using the Q-function
require differential identities which will involve these variables.
Thus, these identities allow one to perform simulations with fermionic
phase space representations. 

We wish to write differential identities for the action of Majorana
operators on the unit-trace Gaussian basis defined in (\ref{eq:MajOpX}).
There is a correspondence between Majorana operators and Fermi operators,
hence we derive four differential identities which are written using
normal, anti-normal ordering and mixed products of normal and anti-normal
ordering. In order to derive the identities for the normalized Gaussian
operators it is convenient to first derive the differential identities
for the unnormalized Majorana operators. The next step is to perform
a change of variables and use the corresponding normalization factor
with its respectively derivative.  The differential identities are
given below and the details of this procedure are explained in~\ref{sec:AppendixMajoranaDiffId},
\ref{sec:AppendixDiffIdNormGO} and \ref{sec:AppendixDiffIdunorderedGO-1}. 

\subsection{Un-normalized Majorana differential identities \label{subsec:Un-normalized-MajDifY}}

In order to derive the differential identities for the normalized
Gaussian operators we need to derive first the corresponding differential
identities for the un-normalized Gaussian operators given in (\ref{eq:GOY}).
These identities are needed in order to obtain the differential identities
in terms of the antisymmetric matrix $\underline{\underline{X}}$.
This matrix is related to the covariance matrix $\underline{\underline{\sigma}}$,
through the expression given in (\ref{eq:XMat}).

We use the convention for matrix derivatives that: 
\begin{equation}
\left[\frac{\rmd}{\rmd\underline{\underline{Y}}}\right]_{\mu\nu}=\frac{\rmd}{\rmd Y_{\nu\mu}}\,.
\end{equation}

These identities are given in terms of the different forms of the
possible orderings between the Majorana variables and the Gaussian
operators and are listed below. The corresponding proofs are shown
in~\ref{sec:AppendixMajoranaDiffId}. 
\begin{itemize}
\item Mixed products:
\begin{equation}
\Biggl\{\hat{\underline{\gamma}}:\hat{\underline{\gamma}}^{T}\hat{\Lambda}^{\left(u\right)}:\Biggr\}=\rmi\left[-\underline{\underline{{\cal I}}}\left(2\underline{\underline{Y}}+\underline{\underline{\mathcal{I}}}\right)\frac{\rmd}{\rmd\underline{\underline{Y}}}+2\underline{\underline{{\cal I}}}\right]\hat{\Lambda}^{\left(u\right)}.\label{eq:MajDiffIdY}
\end{equation}
The explicit form of $\Bigl\{\hat{\underline{\gamma}}:\hat{\underline{\gamma}}^{T}\hat{\Lambda}:\Bigr\}$,
in terms of the creation and annihilation operators is given in~\ref{sec:AppendixFormOrdering}.
\item Normally ordered products:
\begin{equation}
:\hat{\underline{\gamma}}\hat{\underline{\gamma}}^{T}\hat{\Lambda}^{\left(u\right)}:=\rmi\frac{d}{d\underline{\underline{Y}}}\hat{\Lambda}^{\left(u\right)}.\label{eq:UnNormNormalDifId}
\end{equation}
For clarity the detailed expression is given in~\ref{sec:AppendixFormOrdering}.
\item Anti-normally ordered products:
\begin{eqnarray}
\left\{ \underline{\widehat{\gamma}}\widehat{\underline{\gamma}}^{T}\hat{\Lambda}^{\left(u\right)}\right\}  & = & \rmi\underline{\underline{\mathcal{I}}}\left\{ \left(2\underline{\underline{Y}}+\underline{\underline{\mathcal{I}}}\right)\frac{\rmd}{\rmd\underline{\underline{Y}}}\hat{\Lambda}^{u}-2\hat{\Lambda}^{u}\right\} \left(2\underline{\underline{Y}}+\underline{\underline{\mathcal{I}}}\right)\underline{\underline{\mathcal{I}}}.\label{eq:AntiNormDifIdUnGO}
\end{eqnarray}
The full expression is given in~\ref{sec:AppendixFormOrdering}. 
\end{itemize}

\subsection{Normalized Majorana differential identities\label{subsec:Normalized-Majorana-DifId}}

In this section we will give the differential identities for the normalized
Gaussian operators given in (\ref{eq:NormGOX}). We note that in order
to obtain these identities, we need to perform a change of variables
from the matrices $\underline{\underline{Y}}$ to the $\underline{\underline{X}}$
using the relationship given in (\ref{eq:RelXY}) hence:
\begin{equation}
\underline{\underline{Y}}=\left(\underline{\underline{X}}-\underline{\underline{\mathcal{I}}}\right)^{-1}-\underline{\underline{\mathcal{I}}}.\label{eq:ChangeVarYX}
\end{equation}
We also recall that $\hat{\Lambda}\left(\underline{\underline{X}}\right)=N\left(\underline{\underline{X}}\right)\hat{\Lambda}^{\left(u\right)}\left(\underline{\underline{X}}\right),$
where $N\left(\underline{\underline{X}}\right)$ is given in (\ref{eq:NormGOX}),
The normalized differential identities are given below and in \ref{sec:AppendixDiffIdNormGO}
we show the procedure in order to obtain these identities. 
\begin{itemize}
\item Mixed products:
\end{itemize}
\begin{equation}
\Biggl\{\hat{\underline{\gamma}}:\hat{\underline{\gamma}}^{T}\hat{\Lambda}:\Biggr\}=\rmi\left[\underline{\underline{X}}^{+}\frac{\rmd\hat{\Lambda}}{\rmd\underline{\underline{X}}}-\hat{\Lambda}\right]\underline{\underline{X}}^{-}.\label{eq:MajMixProd}
\end{equation}

\begin{equation}
:\hat{\underline{\gamma}}\Biggl\{\hat{\underline{\gamma}}^{T}\hat{\Lambda}\Biggr\}:=\rmi\left[\underline{\underline{X}}^{-}\frac{\rmd\hat{\Lambda}}{\rmd\underline{\underline{X}}}-\hat{\Lambda}\right]\underline{\underline{X}}^{+}.\label{eq:Majmix2}
\end{equation}
\begin{itemize}
\item Normally ordered products:
\begin{equation}
:\hat{\underline{\gamma}}\hat{\underline{\gamma}}^{T}\hat{\Lambda}:=-\rmi\left[\underline{\underline{X}}^{-}\frac{\rmd\hat{\Lambda}}{\rmd\underline{\underline{X}}}-\hat{\Lambda}\right]\underline{\underline{X}}^{-}.\label{eq:NormalNormMajId}
\end{equation}
\item Anti-normally ordered products:
\begin{eqnarray}
\left\{ \underline{\widehat{\gamma}}\widehat{\underline{\gamma}}^{T}\hat{\Lambda}\right\}  & =-\rmi\left[\underline{\underline{X}}^{+}\frac{\rmd\hat{\Lambda}}{\rmd\underline{\underline{X}}}-\hat{\Lambda}\right]\underline{\underline{X}}^{+} & .\label{eq:AntiNormalDifId}
\end{eqnarray}
\end{itemize}

\subsection{Unordered Majorana differential identities\label{subsec:Unordered-Majorana-differential}}

In this section we obtain differential identities that do not use
normal or anti-normal ordering as in the previous section. There is
no natural normal or anti-normal ordering for Majorana operators,
however, the differential identities for the unordered products of
the Majorana operators and the Gaussian basis can be obtained by combining
the four normalized differential identities given in Section \ref{subsec:Normalized-Majorana-DifId}.
The details of the derivation of the differential identities are given
in \ref{sec:AppendixDiffIdunorderedGO-1}. These results are simpler
in the modified variables, $\underline{\underline{x}}=\underline{\underline{{\cal I}}}\underline{\underline{X}}^{T}\underline{\underline{{\cal I}}},$
together with $\underline{\underline{x}}^{\pm}=\underline{\underline{x}}\pm\rmi\underline{\underline{I}}$.
The resulting identities are: 
\begin{itemize}
\item Unordered left product:
\end{itemize}
\begin{equation}
\hat{\underline{\gamma}}\hat{\underline{\gamma}}^{T}\hat{\Lambda}=\rmi\left[\underline{\underline{x}}^{-}\frac{\rmd\hat{\Lambda}}{\rmd\underline{\underline{x}}}\underline{\underline{x}}^{+}-\hat{\Lambda}\underline{\underline{x}}^{+}\right].\label{eq:UnorderedDifId}
\end{equation}
\begin{itemize}
\item Unordered right product:
\end{itemize}
\begin{equation}
\hat{\Lambda}\hat{\underline{\gamma}}\hat{\underline{\gamma}}^{T}=\rmi\left[\underline{\underline{x}}^{+}\frac{\rmd\hat{\Lambda}}{\rmd\underline{\underline{x}}}\underline{\underline{x}}^{-}-\hat{\Lambda}\underline{\underline{x}}^{+}\right].\label{eq:UnorderedDifId-1}
\end{equation}
\begin{itemize}
\item Unordered mixed product:
\end{itemize}
\begin{equation}
\hat{\underline{\gamma}}\hat{\Lambda}\hat{\underline{\gamma}}^{T}=\rmi\left[-\underline{\underline{x}}^{-}\frac{\rmd\hat{\Lambda}}{\rmd\underline{\underline{x}}}\underline{\underline{x}}^{-}+\hat{\Lambda}\underline{\underline{x}}^{-}\right].\label{eq:unmixed}
\end{equation}

\section{Phase space representations\label{sec:Phase-space-representations}}

Phase space representations have been widely used in quantum optics
and ultracold atoms. A fermionic $P$-representation can be introduced
using Grassmann variables, which are variables that anti-commute~\cite{Cahill_Glauber_fermions_1999},
but these are exponentially complex and hence mostly useful for analytic
calculations, although they can be used to obtain complex covariance
type equations~\cite{dalton2016grassmann}.

Another approach is to define fermionic phase space using the Gaussian
operators. For fermions there are two known complete representations,
which are the fermionic P-representation~\cite{Cahill_Glauber_fermions_1999,Corney_PD_PRL2004_GQMC_ferm_bos,Corney_PD_JPA_2006_GR_fermions,Corney_PD_PRB_2006_GPSR_fermions,Corboz_Chapter_PhaseSpaceMethodsFermions}
and the fermionic Q-function \cite{FermiQ}. In this case the phase
space variables correspond to the matrix of the Gaussian operator,
which are the elements of the covariance matrix. Using the Majorana
Gaussian representation it is also possible to define phase space
representations in terms of the Majorana variables and an antisymmetric
matrix. These are the analogous to the fermionic phase space representations.

\subsection{Majorana P-functions \label{subsec:Majorana-P-representation}}

A Majorana fermionic P-function can be defined in analogy to the fermionic
P-function using raising and lowering operators~\cite{Cahill_Glauber_fermions_1999,Corney_PD_PRL2004_GQMC_ferm_bos,Corney_PD_JPA_2006_GR_fermions,Corney_PD_PRB_2006_GPSR_fermions,Corboz_Chapter_PhaseSpaceMethodsFermions}.
In this case the density matrix $\hat{\rho}(t)$ is expanded in terms
of the Gaussian basis given in (\ref{eq:MajOpX}). The phase space
variables are either the antisymmetric matrix $\underline{\underline{X}}$,
or its cousin $\underline{\underline{x}}$. Each may be more suitable,
depending on whether one wishes to use ordered or unordered Majorana
identities. Here we choose the unordered form: 
\[
\hat{\rho}(\tau)=\int P\left(\underline{\underline{x}},\tau\right)\hat{\Lambda}\left(\underline{\underline{x}}\right)\prod_{i}\rmd x_{i}.
\]
The $P(\underline{\underline{x}},\tau)$ distribution has the same
properties as the fermionic $P$-representation:
\begin{enumerate}
\item \textbf{Normalization}: It is normalized to one:
\[
\int_{{\cal D}_{C}}P\left(\underline{\underline{x}}\right)\prod_{i}\rmd x_{i}=1.
\]
\item \textbf{Moments}: Expectation values of observables $O$ are obtained
as:
\begin{eqnarray*}
\left\langle O\right\rangle  & = & \int_{{\cal D}_{C}}P\left(\underline{\underline{x}},\tau\right)\Tr\left[O\hat{\Lambda}\left(\underline{\underline{x}}\right)\right]\prod_{i}\rmd x_{i}\equiv\left\langle O\right\rangle _{P}.
\end{eqnarray*}
\end{enumerate}

\subsection{Majorana Q-functions}

A Majorana Q-function can be defined as:
\begin{equation}
Q\left(\underline{\underline{x}}\right)=\Tr\left[\hat{\rho}\hat{\Lambda}^{N}\left(\underline{\underline{x}}\right)\right].\label{eq:MajQf}
\end{equation}
This Majorana Q-function has the same properties as the fermionic
Q-function which are:
\begin{enumerate}
\item It is defined for any quantum density-matrix. 
\item It is a positive probability distribution.
\item Observables are moments of the distribution. These are obtained as:
\begin{eqnarray}
\left\langle \hat{O}_{n}\right\rangle  & = & \int_{{\cal D}_{R}}Q\left(\underline{\underline{x}}\right)O_{n}\left(\underline{\underline{x}}\right)\prod_{i}\rmd x_{i}=\left\langle \hat{O}_{n}\left(\left(\underline{\underline{x}}\right)\right)\right\rangle _{Q}.\label{eq:moment}
\end{eqnarray}
\end{enumerate}

\subsection{Single mode Majorana Q-function}

It is instructive to have an expression for the Q function in the
single mode case. Utilizing \ref{eq:MajOpX}, where there is only
one real variable, $x\equiv x_{12}$, and:
\begin{equation}
\underline{\underline{x}}=\left[\begin{array}{cc}
0 & x\\
-x & 0
\end{array}\right].\label{eq:U0-1}
\end{equation}
We can begin with the expression for the single mode Gaussian operator
in Majorana form, which can be re- expressed using the result that:
\begin{equation}
\widehat{n}=\frac{1+\rmi\hat{\gamma}_{1}\hat{\gamma}_{2}}{2}
\end{equation}
so that 
\begin{equation}
\hat{\Lambda}\Biggl(x\Biggr)=\frac{1}{2}+\frac{\rmi}{2}\hat{\gamma}_{1}\hat{\gamma}_{2}x.
\end{equation}

All physical density operators are also Gaussian operators in single
mode case\cite{Corney_PD_JPA_2006_GR_fermions}, since if $n\equiv\left\langle \hat{n}\right\rangle $,
the most general density matrix is

\begin{equation}
\hat{\rho}=\hat{\Lambda}_{1}\left(n\right)=\left(1-n\right)\left|0\right\rangle \left\langle 0\right|+n\left|1\right\rangle \left\langle 1\right|.
\end{equation}
We arrive at an expression for Majorana Q-function by using the above
two equations in \ref{eq:MajQf}, to give:

\begin{equation}
Q\left(x\right)=\frac{1}{{\cal N}}S\left(x\right)\left[\left(\frac{1-x}{2}\right)+nx\right],
\end{equation}
 where $S\left(x\right)=\left[1+x^{2}\right]^{2p}$. 

Just as with the unit trace Gaussian, this has a symmetry between
particles and holes, since changing the occupation from $n$ to $1-n$
simply changes the sign of the argument $x$. We note that in the
absence of the normalization factor $S$, the distribution $Q\left(x\right)$
would have a discontinuity at the boundary.

\subsection{Observables}

We can obtain observables in terms of the Majorana Q-function using
the Majorana differential identities in (\ref{eq:moment}). As an
example of this, we consider the Majorana correlation function,
\begin{equation}
\hat{X}_{\mu\nu}\equiv\frac{\rmi}{2}\left[\gamma_{\mu},\gamma_{\nu}\right].\label{eq:Xcap}
\end{equation}
This is a general hermitian observable, which includes the occupation
number operators since:
\begin{equation}
\widehat{n}_{i}=\frac{1+\hat{X}_{i,M+i}}{2}\,.
\end{equation}
The expectation is defined as:
\begin{equation}
\Biggl\langle\hat{X}_{\mu\nu}\Biggr\rangle=Tr\left[\hat{\rho}\widehat{X}_{\mu\upsilon}\right].\label{eq:mom1}
\end{equation}
Next, using the resolution of identity and the unordered product ordering,
we can arrive at the expression:
\begin{equation}
\Biggl\langle\hat{X}_{\mu\nu}\Biggr\rangle=\frac{\rmi}{2}\int\Tr\left[\widehat{\rho}\left[\widehat{\gamma}_{\mu}\widehat{\gamma}_{\upsilon}\hat{\Lambda}^{N}-\widehat{\gamma}_{\upsilon}\widehat{\gamma}_{\mu}\hat{\Lambda}^{N}\right]\right]\rmd\underline{\underline{x}}.\label{eq:mom2}
\end{equation}
Now we can apply the differential identities corresponding to the
unordered product given in (\ref{eq:UnorderedDifId}) obtaining:
\begin{eqnarray}
\Bigl\langle\widehat{X}_{\mu\upsilon}\Bigr\rangle & = & \frac{1}{2}\int\Tr\left[\widehat{\rho}\frac{S}{{\cal N}}\left[-x_{\mu\alpha}^{-}\frac{\rmd\hat{\Lambda}}{\rmd x_{\beta\alpha}}x_{\beta\upsilon}^{+}+\hat{\Lambda}x_{\mu\upsilon}^{+}\right]\rmd\underline{\underline{x}}\right]\nonumber \\
 &  & -\frac{1}{2}\int\Tr\left[\widehat{\rho}\frac{S}{{\cal N}}\left[-x_{\upsilon\alpha}^{-}\frac{\rmd\hat{\Lambda}}{\rmd x_{\beta\alpha}}x_{\beta\mu}^{+}+\hat{\Lambda}x_{\upsilon\mu}^{+}\right]\rmd\underline{\underline{x}}\right].
\end{eqnarray}
Considering the limit $S\rightarrow1$ and assuming that the boundary
terms vanish and also using the definition of the Q-function, the
above equation can be written as:
\begin{equation}
\fl\Bigl\langle\widehat{X}_{\mu\upsilon}\Bigr\rangle=\frac{1}{2}\int\left[-x_{\mu\alpha}^{-}\frac{\rmd}{\rmd x_{\beta\alpha}}x_{\beta\upsilon}^{+}+x_{\mu\upsilon}^{+}\right]Q\rmd\underline{\underline{x}}-\frac{1}{2}\int\left[-x_{\upsilon\alpha}^{-}\frac{\rmd}{\rmd x_{\beta\alpha}}x_{\beta\mu}^{+}+x_{\upsilon\mu}^{+}\right]Q\rmd\underline{\underline{x}}.
\end{equation}
Using the the chain rule the above equation is:

\begin{eqnarray}
\Bigl\langle\widehat{X}_{\mu\upsilon}\Bigr\rangle & = & -\frac{1}{2}\int\left[\frac{\rmd}{\rmd x_{\beta\alpha}}\left[x_{\mu\alpha}^{-}x_{\beta\upsilon}^{+}\right]-2x_{\mu\upsilon}\left(2M-1\right)-x_{\mu\upsilon}^{+}\right]Q\rmd\underline{\underline{x}}\nonumber \\
 &  & +\frac{1}{2}\int\left[\frac{\rmd}{\rmd x_{\beta\alpha}}\left[x_{\upsilon\alpha}^{-}x_{\beta\mu}^{+}\right]-2x_{\upsilon\mu}\left(2M-1\right)-x_{\upsilon\mu}^{-}\right]Q\rmd\underline{\underline{x}}
\end{eqnarray}

Here we have assumed that the boundary terms for the normal components
of $x_{\mu\alpha}^{-}x_{\beta\upsilon}^{+}Q$ vanish at the classical
domain boundary, so that the corresponding total derivative integrates
to zero:
\begin{equation}
\int\frac{\rmd}{\rmd x_{\beta\alpha}}\left[x_{\mu\alpha}^{-}x_{\beta\upsilon}^{+}Q\right]\rmd\underline{\underline{x}}=0\,.
\end{equation}
This will leads to the required observable equation:
\begin{equation}
\Bigl\langle\widehat{\underline{\underline{X}}}\Bigr\rangle=\left(4M-1\right)\int\underline{\underline{x}}Q\left(\underline{\underline{x}}\right)\rmd\underline{\underline{x}}.
\end{equation}
Similarly, we can extend this method to evaluate higher order correlations
as well. The assumption of vanishing boundary terms is an important
restriction on our results, and would need to be checked in individual
cases.

\section{Hamiltonian and time evolution \label{sec:Time-Evolution-MQf}}

Majorana variables have been used to study decoherence effects in
Fermi systems~\cite{Prosen:2008} or fermionic entanglement~\cite{Eisler:2015,Meichanetzidis:2016Ent}.
These effects are related to the dynamical evolution of the systems.
The Majorana differential identities are useful in order to perform
dynamical simulations. In order to illustrate this, we will consider
the simplest case of quadratic Hamiltonians. While higher order terms
can be treated, these result in diffusion-type differential operators,
which are outside the scope of the present article. 

\subsection{Quadratic Hamiltonians}

The symmetries of fermionic covariance matrices and quadratic Hamiltonians
have identical properties. The general hermitian quadratic Hamiltonian
has been widely investigated, and has the form:
\begin{equation}
\hat{H}=\frac{1}{2}\left[\hat{\bm{a}}^{\dagger}\mathbf{h}\hat{\bm{a}}-\hat{\bm{a}}\bm{h}^{T}\hat{\bm{a}}^{\dagger}+\hat{\bm{a}}^{\dagger}\bm{\Delta}\hat{\bm{a}}^{\dagger}-\hat{\bm{a}}\bm{\Delta}^{*}\hat{\bm{a}}\right].\label{eq:BdGHam-2}
\end{equation}
This Hamiltonian is also known as the Bogoliubov- de Gennes Hamiltonian,
which can be written in matrix form using the extended ladder operators
as:
\begin{equation}
\hat{H}=\frac{1}{2}\hat{\underline{a}}^{\dagger}\underline{\underline{H}}\hat{\underline{a}},\label{eq:BdGHamMat-1}
\end{equation}
where the matrix $\underline{\underline{H}}$ is defined as:

\begin{equation}
\underline{\underline{H}}=\left(\begin{array}{cc}
\mathbf{h} & \bm{\Delta}\\
-\bm{\Delta^{*}} & -\mathbf{h}^{T}
\end{array}\right),\label{eq:H_Matrix-2}
\end{equation}
with $\mathbf{h}=\mathbf{h}^{\dagger}$ and $\bm{\Delta}=-\bm{\Delta}^{T}$
as the only symmetry restrictions. Altland and Zirnbauer~\cite{Altland_Zirnbauer:1997}
consider this Hamiltonian in order to define the non-standard symmetry
classes. This is done by considering how imposing time-reversal and/or
spin-rotation symmetry on the Hamiltonian leads to different types
of symmetry classes. Each one of these corresponds to a particular
type of Lie group that has a corresponding mathematical symmetric
space. 

Here we will focus on class D symmetry, which has neither time-reversal
nor spin-rotation invariance, allowing us to treat an arbitrary Fermi
system, which means that $\underline{\underline{H}}=\underline{\underline{H}}^{\dagger}.$
Using the relationship for the ladder operators and the Majorana operators
given in (\ref{eq:OpMajUn}) we can express the Hamiltonian of (\ref{eq:BdGHamMat-1})
in the following form~\cite{Kitaev:2009,Alicea:2012}:
\begin{equation}
\hat{H}=\frac{\rmi\hbar}{2}\hat{\underline{\gamma}}^{T}\underline{\underline{\Omega}}\hat{\underline{\gamma}}=\frac{\hbar}{2}\Omega_{\mu\nu}\hat{X}_{\mu\nu}.\label{eq:BdGHam_MajOp-2}
\end{equation}
The expression of the matrix $\underline{\underline{\Omega}}$ in
terms of matrices $\mathbf{h}$ and $\bm{\Delta}$ is:
\begin{equation}
\underline{\underline{\Omega}}=\frac{1}{2\rmi\hbar}\left(\begin{array}{cc}
\mathbf{h}_{-}+\bm{\Delta}_{-} & i\mathbf{h}_{+}-i\bm{\Delta}_{+}\\
-i\mathbf{h}_{+}-i\bm{\Delta}_{+} & \mathbf{h}_{-}-\bm{\Delta}_{-}
\end{array}\right)\,.\label{eq:HamMatAntiSym-1}
\end{equation}
Here we have defined $\mathbf{h}_{\pm}=\mathbf{h}\pm\mathbf{h}^{T}$
and $\bm{\Delta}_{\pm}=\bm{\Delta}\pm\bm{\Delta}^{*}$. We note that
$\underline{\underline{\Omega}}$ is a real anti-symmetric matrix:
$\underline{\underline{\Omega}}^{T}=-\underline{\underline{\Omega}}$,
and therefore satisfies the group properties of these matrices, with
a well-defined Haar measure.

\subsection{Time evolution}

The time evolution equation of the Majorana Q-function is obtained
by considering that:
\begin{equation}
\frac{\rmd Q\left(\underline{\underline{x}}\right)}{\rmd t}=\Tr\left[\frac{\rmd\hat{\rho}}{\rmd t}\hat{\Lambda}^{N}\left(\underline{\underline{x}}\right)\right].\label{eq:TE_MQf}
\end{equation}
Here we have used the definition of the Majorana Q-function given
in (\ref{eq:MajQf}). Next we consider the time-evolution equation
for the density operator:
\begin{eqnarray}
i\hbar\frac{\partial}{\partial t}\hat{\rho} & = & \left[\hat{H},\hat{\rho}\right].\label{eq:TimeEvolutionDM}
\end{eqnarray}
On substituting this on the time-evolution equation for the Majorana
Q-function and using the cyclic properties of the trace, we get:
\begin{equation}
\frac{\rmd Q\left(\underline{\underline{x}}\right)}{\rmd t}=\frac{1}{\rmi\hbar}\Tr\left[\left[\hat{\Lambda}^{N}\left(\underline{\underline{x}}\right),\hat{H}\right]\hat{\rho}\right].\label{eq:TT}
\end{equation}
Here we will consider the general Hamiltonian $\hat{H}$ given in
(\ref{eq:BdGHam_MajOp-2}). Therefore we get that the time-evolution
equation for the Majorana Q-function is:
\[
\frac{\rmd Q\left(\underline{\underline{x}}\right)}{\rmd t}=\frac{1}{2\rmi}\Tr\left[\hat{\Lambda}^{N}\Omega_{\mu\nu}\hat{X}_{\mu\nu}\hat{\rho}-\Omega_{\mu\nu}\hat{X}_{\mu\nu}\hat{\Lambda}^{N}\hat{\rho}\right].
\]
We now use the definition of $\underline{\underline{\hat{X}}}$ given
in (\ref{eq:Xcap}), obtaining:
\begin{equation}
\frac{\rmd Q\left(\underline{\underline{x}}\right)}{\rmd t}=-\frac{1}{4}\Tr\left[\Omega_{\mu\nu}\left[\gamma_{\mu}\gamma_{\nu}-\gamma_{\nu}\gamma_{\mu},\hat{\Lambda}^{N}\right]\hat{\rho}\right].\label{eq:TF}
\end{equation}

The differential identity for $\left[\gamma_{\mu}\gamma_{\nu}-\gamma_{\nu}\gamma_{\mu},\widehat{\Lambda}\right]$
is derived from (\ref{eq:UnorderedDifId}). This is given below:
\begin{equation}
\left[\gamma_{\mu}\gamma_{\nu}-\gamma_{\nu}\gamma_{\mu},\widehat{\Lambda}\right]=4\left[x_{\kappa\upsilon}\frac{\rmd\hat{\Lambda}}{\rmd x_{\kappa\mu}}-x_{\mu\kappa}\frac{\rmd\hat{\Lambda}}{\rmd x_{\upsilon\kappa}}\right].\label{eq:COMMU}
\end{equation}
On substituting (\ref{eq:COMMU}) in (\ref{eq:TF}), we get:

\begin{equation}
\frac{\rmd Q\left(\underline{\underline{X}}\right)}{\rmd t}=-\Tr\left[\Omega_{\mu\nu}\frac{S}{{\cal N}}\left[-x_{\mu\kappa}\frac{\rmd\hat{\Lambda}}{\rmd x_{\upsilon\kappa}}+\frac{\rmd\hat{\Lambda}}{\rmd x_{\kappa\mu}}x_{\kappa\upsilon}\right]\hat{\rho}\right].\label{eq:dQ/dt}
\end{equation}
Following the steps outlined in \ref{sec:AppendixTimeevolution},
we get:
\begin{equation}
\frac{\rmd Q\left(\underline{\underline{X}}\right)}{\rmd t}=-\Omega_{\mu\nu}\left[-x_{\mu\kappa}\frac{\rmd Q}{\rmd x_{\upsilon\kappa}}+\frac{\rmd Q}{\rmd x_{\kappa\mu}}x_{\kappa\upsilon}\right]\label{eq:TEf}
\end{equation}

Next, using the product rule, (\ref{eq:TEf}) can be written as:
\begin{eqnarray}
\frac{\rmd Q\left(\underline{\underline{X}}\right)}{\rmd t} & = & \Omega_{\mu\nu}\left[\frac{\rmd}{\rmd x_{\upsilon\kappa}}\left(x_{\mu\kappa}Q\right)-\frac{\rmd}{\rmd x_{\kappa\mu}}\left(x_{\kappa\upsilon}Q\right)\right].\label{eq:TEf2}
\end{eqnarray}
The method of characteristics allows us to solve the above equation
as:
\begin{equation}
\frac{d\underline{\underline{x}}}{dt}=\left[\underline{\underline{\Omega}},\underline{\underline{x}}\right].\label{eq:TEX}
\end{equation}

We note that the Heisenberg equations of motion for the Majorana operators
in this case are identical, and are given by:
\begin{equation}
\frac{\rmd\underline{\underline{\hat{X}}}\left(t\right)}{\rmd t}=\left[\underline{\underline{\Omega}},\underline{\hat{\underline{X}}}\right].
\end{equation}
This result is valid for completely arbitrary quadratic real antisymmetric
matrices $\underline{\underline{\Omega}}$, even with anomalous terms
or complex frequency matrices. Since the operator equations for $\underline{\underline{\hat{X}}}$
are simply proportional to the mean value equations, which in turn
are proportional to the characteristic equations for $\underline{\underline{x}}$,
this provides a verification of the phase-space results. The utility
of the Q-function method in this case is that the general Q-function
can be computed given an arbitrary initial distribution. One is not
restricted to just calculating mean values of the lowest order correlation
function.

\subsection{Bosonic Q-function}

For comparison purposes, we can also write a time evolution for the
bosonic Q-function \cite{Husimi1940} in a similar form that the one
described in the above section for fermions. In this case we consider
the following Hamiltonian of a non-interacting Bose gas:

\begin{equation}
\hat{H}=\hbar\hat{\bm{a}}^{\dagger}\mathbf{\bm{\omega}}\hat{\bm{a}}.
\end{equation}

Following the standard techniques for phase-space representations,
where one uses the corresponding identities that map phase space variables
into c-numbers \cite{Drummond:2014BookQT}, we obtain the following
time evolution equation for the bosonic Q function, $Q_{B}$:
\begin{equation}
\frac{dQ_{B}(\boldsymbol{\alpha})}{dt}=i\bm{\bm{\omega}}\left[\frac{\partial}{\partial\boldsymbol{\alpha}}\boldsymbol{\alpha}-\frac{\partial}{\partial\boldsymbol{\alpha}^{*}}\boldsymbol{\alpha}^{*}\right]Q_{B}.\label{eq:btimeev}
\end{equation}
We use the methods of characteristics in order to solve this differential
equation obtaining:

\begin{equation}
\frac{d\boldsymbol{\alpha}}{dt}=-i\bm{\bm{\omega}}\boldsymbol{\alpha}.\label{eq:char}
\end{equation}
Next, we define the following two real vectors:
\begin{eqnarray*}
\boldsymbol{\alpha}_{x} & = & \frac{1}{2}\left[\boldsymbol{\alpha}+\boldsymbol{\alpha}^{*}\right],\\
\boldsymbol{\alpha}_{y} & = & \frac{1}{2i}\left[\boldsymbol{\alpha}-\boldsymbol{\alpha}^{*}\right].
\end{eqnarray*}

Consequently we can introduce the following real vector:
\begin{equation}
\underline{\boldsymbol{\alpha}}=\left[\begin{array}{c}
\boldsymbol{\alpha}_{x}\\
\boldsymbol{\alpha}_{y}
\end{array}\right],\label{eq:quad}
\end{equation}
and a matrix of quadrature correlations, analogous to the Majorana
phase space matrix, where 
\[
\underline{\underline{x}}_{b}=\underline{x}\underline{x}^{T}.
\]
Using these expressions in (\ref{eq:char}), leads to a time evolution
equation for $\underline{\underline{x}}_{b}$ of the form:
\begin{equation}
\frac{d}{dt}\left(\underline{\underline{x}}_{b}\right)=\left[\underline{\underline{\Omega}}_{b},\underline{\underline{x}}_{b}\right],\label{eq:Xb}
\end{equation}
where $\underline{\underline{\Omega}}_{b}=\left(\begin{array}{cc}
0 & \boldsymbol{\omega}\\
-\boldsymbol{\mathbf{\omega}} & 0
\end{array}\right).$ 

In summary, for the usual bosonic case, one also obtains a simple
characteristic evolution, provided the Hamiltonian has no anomalous
terms. More generally, the results are more complicated, and do not
follow a simple first-order equation. In the Fermi Q-function case,
a first-order characteristic evolution is obtained for any quadratic
Hamiltonian.

\section{Time-Evolution of open quantum systems\label{sec:TimeEvOpenSyst}}

We now consider the following application of our method: the time
evolution of an open quantum system. This will include the interaction
of a system with the environment, which will show that our method
can treat dissipative systems as well. The system is a small quantum
dot coupled to a zero temperature reservoir. Here it is convenient
to use the ordered identities, as the dissipative master-equation
terms have normal ordering.

\subsection{Master Equation}

The time evolution of the density operator of this model is given
by a master equation~\cite{Corney_PD_PRB_2006_GPSR_fermions}:

\begin{equation}
\frac{d\hat{\rho}}{dt}=-i\omega\hat{a}^{\dagger}\hat{a}\hat{\rho}+i\omega\hat{\rho}\hat{a}^{\dagger}\hat{a}+\gamma\left(\hat{a}\hat{\rho}\hat{a}^{\dagger}-\frac{1}{2}\hat{a}^{\dagger}\hat{a}\hat{\rho}-\frac{1}{2}\hat{\rho}\hat{a}^{\dagger}\hat{a}\right).\label{eq:DensityOpQD}
\end{equation}
In the case of a multi-mode quantum system we get
\begin{equation}
\frac{d\hat{\rho}}{dt}=-i\omega_{ji}\hat{a}_{i}^{\dagger}\hat{a}_{j}\hat{\rho}+i\omega_{ji}\hat{\rho}\hat{a}_{i}^{\dagger}\hat{a}_{j}+\gamma_{ij}\left(\hat{a}_{i}\hat{\rho}\hat{a}_{j}^{\dagger}-\frac{1}{2}\hat{a}_{j}^{\dagger}\hat{a}_{i}\hat{\rho}-\frac{1}{2}\hat{\rho}\hat{a}_{j}^{\dagger}\hat{a}_{i}\right),\label{eq:Denmul}
\end{equation}
provided $\bm{\omega}=\bm{\omega}^{T}$ and $\bm{\gamma}=\bm{\gamma}^{T}$.
On substituting this equation for the time evolution of the Q-function
given in (\ref{eq:TE_MQf}) we obtain: 

\begin{eqnarray}
\frac{dQ}{dt} & = & i\Tr\Biggl[\omega_{ji}\left[\widehat{a}_{i}^{\dagger}\widehat{a}_{j},\hat{\Lambda}^{N}\left(\underline{\underline{X}}\right)\right]\widehat{\rho}\Biggr]+\Tr\left[\hat{\Lambda}^{N}(\underline{\underline{X}})\gamma_{ij}\widehat{a}_{i}\widehat{\rho}\widehat{a}_{j}^{\dagger}\right]\nonumber \\
 &  & -\frac{1}{2}\Tr\left[\gamma_{ij}\left[\widehat{a}_{i}^{\dagger}\widehat{a}_{j},\hat{\Lambda}^{N}\left(\underline{\underline{X}}\right)\right]_{+}\widehat{\rho}\right].\label{eq:timeevo1}
\end{eqnarray}
We now wish to express the above results in terms of the Majorana
differential identities given in Section \ref{sec:Majorana-differential-properties}.
First, using the expressions of the different orderings of the Majorana
variables and the Gaussian operators given in \ref{sec:AppendixDiffIdNormGO}
we rewrite (\ref{eq:timeevo1}) as:

\begin{eqnarray}
\frac{dQ}{dt} & = & -\frac{i}{2}\frac{1}{{\cal N}}S\Tr\left[\tilde{\Omega}_{\kappa\nu}\Bigl\{\hat{\underline{\gamma}}:\hat{\underline{\gamma}}^{T}\hat{\Lambda}:\Bigr\}_{\nu\kappa}\widehat{\rho}\right]\nonumber \\
 &  & -\frac{1}{2i}\frac{1}{{\cal N}}S\Tr\left[\Upsilon_{\kappa\nu}\left(:\hat{\underline{\gamma}}\hat{\underline{\gamma}}^{T}\hat{\Lambda}:{}_{\nu\kappa}-\Bigl\{\hat{\underline{\gamma}}:\hat{\underline{\gamma}}^{T}\hat{\Lambda}:\Bigr\}_{\nu\kappa}\right)\widehat{\rho}\right]-\gamma_{ij}\delta_{ij}Q.\label{eq:timeevo2}
\end{eqnarray}
Here we have defined 

\begin{eqnarray}
\underline{\underline{\tilde{\Omega}}} & = & \left(\begin{array}{cc}
\bm{\omega} & \mathbf{0}\\
\mathbf{0} & \bm{\omega}
\end{array}\right),\qquad{\rm and}\label{eq:OmegaQDot}\\
\underline{\underline{\Upsilon}} & = & \left(\begin{array}{cc}
\mathbf{0} & -\frac{\bm{\gamma}}{2}\\
\bm{\frac{\gamma}{2}} & \mathbf{0}
\end{array}\right).\label{eq:Gamma}
\end{eqnarray}
On using the Majorana differential identities given in (\ref{eq:MajMixProd})
and (\ref{eq:NormalNormMajId}) we get that the time evolution equation
is:

\begin{eqnarray}
\frac{dQ}{dt} & = & \frac{1}{2}\tilde{\Omega}_{\kappa\nu}\frac{d}{dX_{p\ell}}\left(X_{\nu\ell}^{+}QX_{p\kappa}^{-}\right)+\Upsilon_{\kappa\nu}\frac{d}{dX_{pl}}\left(X_{\nu\ell}QX_{p\kappa}^{-}\right)\nonumber \\
 &  & -\left(4M-1\right)X_{\nu\kappa}\Upsilon_{\kappa\nu}Q+\left(2M-1\right)\Upsilon_{\kappa\nu}\mathcal{I}_{\nu\kappa}.\label{eq:finalEx}
\end{eqnarray}
Details of the calculations are given in~\ref{sec:AppTimeEvOpenQS}.
We will next illustrate the dynamic behavior for this dissipative
system for the single mode case. 

\subsection{Single mode case}

We wish to show the dynamic behavior of the open quantum system. In
order to do this, we will study the nature of trajectories for the
single mode case, where $\underline{\underline{X}}=\underline{\underline{x}}$.
In this case, defining $X=X_{12}$, (\ref{eq:finalEx}) reduces to: 

\begin{equation}
\frac{dQ}{dt}=\gamma\left[\frac{d}{dX}\left[QX\left(X-1\right)\right]\right]+\gamma\left(1-3X\right)Q.\label{eq:TevQ_QDot_SingleMode}
\end{equation}
We can solve the above differential equation using method of characteristics
which leads to:

\begin{equation}
\frac{dX}{dt}=\gamma X(1-X).\label{eq:dXdtQdotSM}
\end{equation}
On integrating this equation we get:

\begin{equation}
t-t_{0}=\ln\left[\frac{X}{1-X}\right].
\end{equation}
This can be written as:

\begin{equation}
e^{-(t-t_{0})}=\frac{1}{X}-1,
\end{equation}
which gives two possible solutions:

\begin{equation}
X=\frac{1}{1\pm e^{-(t-t_{0})}}.
\end{equation}

\begin{figure}[H]
\begin{centering}
\includegraphics[scale=0.25]{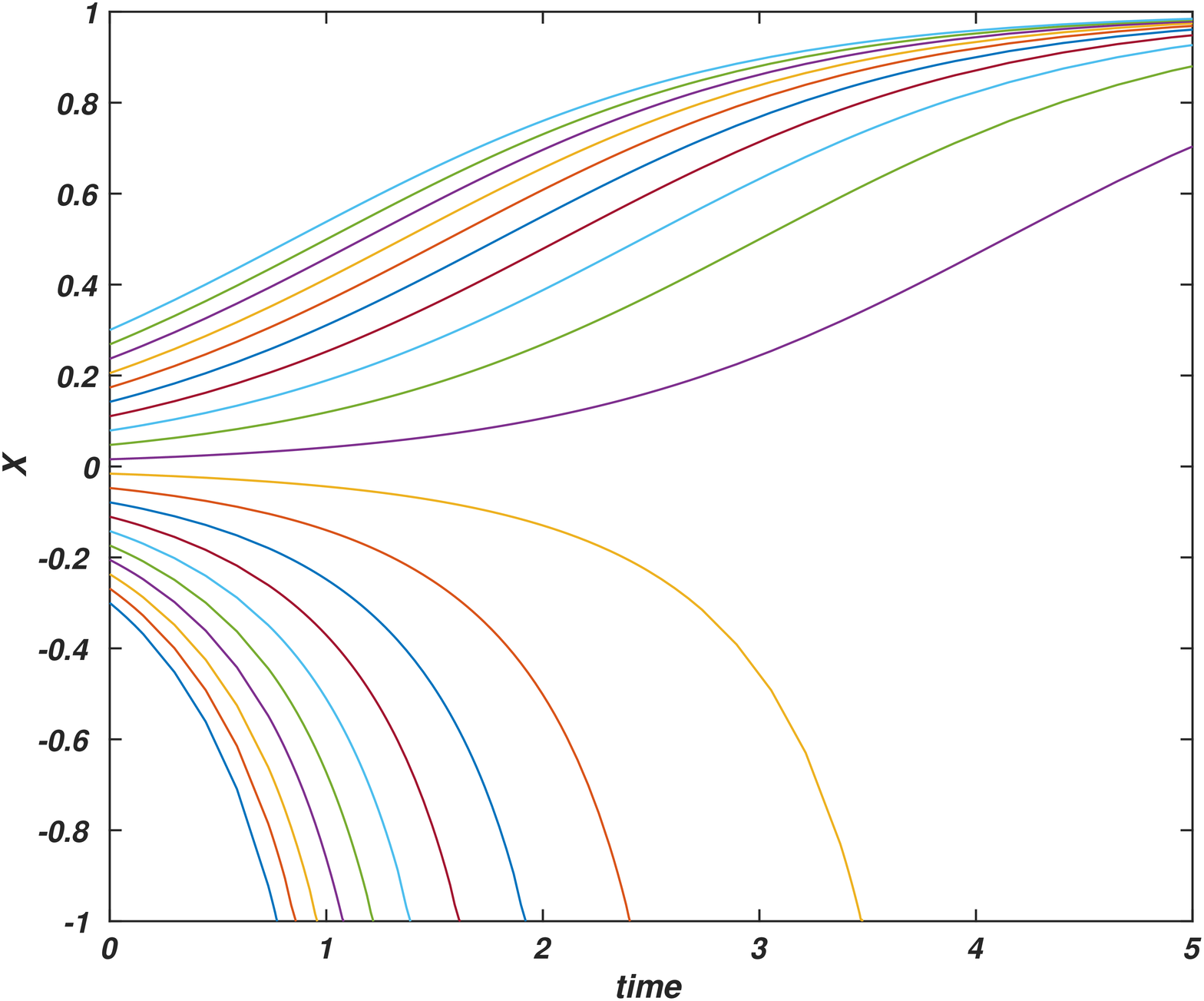}
\par\end{centering}
\caption{Dynamics of a single mode quantum dot coupled to a zero temperature
reservoir.\label{fig:DynQD} }
\end{figure}

In order to get this result we have used that since $t_{0}$ is arbitrary,
it can be chosen as complex, so for $X<0$, one has: $t_{0}\rightarrow i\pi+t_{0}$,
which gives the minus sign. Therefore, if $t\rightarrow\infty$, then,
either with the plus sign, $X(0)>0$, leads to $X\rightarrow1$ or
with the minus sign, $X(0)<0$, leads to $X\rightarrow-\infty.$ This
means that for $X(0)<0$ the trajectory crosses the boundary after
a finite time, and is lost. The solution of the differential equation
(\ref{eq:TevQ_QDot_SingleMode}) is depicted graphically in figure
\ref{fig:DynQD}. 

One usually has a first order Fokker Plank equation when dealing with
a damped P-function for bosons. It decays down to a delta function.
This means that for the bosonic P function, every trajectory decays
to zero. Then the long-time solution in that case is a delta function
with zero width. However, for bosons, the Q-function method would
include a diffusion term, and give a finite width.

In our case we have a first-order differential equation for a fermionic
Q function. In this case we get a finite width for the ground state
solution for the single mode case. We would normally expect to obtain
this through a diffusion term in the Fokker Planck equation. However
we have no diffusion term in our calculation, but we still get a finite
width, which makes the dynamics unusual.

In this case, the Q-function dissipative dynamics has both sources
and sinks, with weighted trajectories being generated at the origin,
and being lost at the boundary at $X=-1$.

\section{Summary\label{sec:Summary_Conclusion}}

In summary, we have introduced a formalism for a fermionic phase space
representation in terms of Majorana operators and a phase-space of
antisymmetric matrices defined within a Cartan bounded homogeneous
domain. This phase space representation uses as a basis the Majorana
Gaussian operators, whose symmetry group is associated with the same
classical domain of real antisymmetric matrices. 

We have derived differential identities for the Majorana Gaussian
operators. These identities are relevant to the important tasks of
computing observables and carrying out dynamical simulations. As
a simple illustration, we have derived time evolution equations of
the Majorana Q-function for the number conserving Hamiltonian, in
the case of a general quadratic Hamiltonian. 

We have also obtained time evolution equations for a quantum dot coupled
to a zero temperature reservoir, which is an example of a dissipative
system. The results obtained are quite different to those in the bosonic
Q-function representation of Husimi. For example, we can treat all
quadratic Hamiltonians as simple trajectory evolution - not just those
with number-conserving Hamiltonians.

These results show the usefulness of the Majorana differential identities.
As an example, our results can also be used to study the dynamics
of shock-wave formation in a one-dimensional Fermi gas at finite temperature.
This study will be carried out elsewhere.

\ack{}{We acknowledge the support of the Australian Research Council.}

\appendix

\section{Explicit form of the different orderings for the Majorana variables
and Gaussian operators\label{sec:AppendixFormOrdering}}

In this section we write the explicit forms of the possible different
orderings of the Majorana variables with the Majorana Gaussian operators.
These expressions are given following the order in which the differential
identities were given. 

\subsection{First mixed product\label{subsec:Mixed-product1}}

\begin{eqnarray}
 &  & \fl\Bigl\{\hat{\underline{\gamma}}:\hat{\underline{\gamma}}^{T}\hat{\Lambda}:\Bigr\}=\nonumber \\
 &  & \fl\left(\begin{array}{cc}
\bm{T}_{1}-\left(\hat{\Lambda}\hat{\bm{a}}\hat{\bm{a}}^{\dagger}\right)^{T}-\left(\hat{\bm{a}}^{\dagger T}\hat{\Lambda}\hat{\bm{a}}^{\dagger}\right)^{T} & -\rmi\left[\bm{T}_{2}-\left(\hat{\Lambda}\hat{\bm{a}}\hat{\bm{a}}^{\dagger}\right)^{T}+\left(\hat{\bm{a}}^{\dagger T}\hat{\Lambda}\hat{\bm{a}}^{\dagger}\right)^{T}\right]\\
-\rmi\left[\bm{T}_{1}+\left(\hat{\Lambda}\hat{\bm{a}}\hat{\bm{a}}^{\dagger}\right)^{T}+\left(\hat{\bm{a}}^{\dagger T}\hat{\Lambda}\hat{\bm{a}}^{\dagger}\right)^{T}\right] & -\left[\bm{T}_{2}+\left(\hat{\bm{a}}^{\dagger T}\hat{\Lambda}\hat{\bm{a}}^{\dagger}\right)^{T}-\left(\hat{\Lambda}\hat{\bm{a}}\hat{\bm{a}}^{\dagger}\right)^{T}\right]
\end{array}\right),\label{eq:ExpFMP}
\end{eqnarray}
where $\bm{T}_{1}=\hat{\bm{a}}\hat{\Lambda}\hat{\bm{a}}^{T}+\hat{\bm{a}}\hat{\bm{a}}^{\dagger}\hat{\Lambda}$
and $\bm{T}_{2}=\hat{\bm{a}}\hat{\Lambda}\hat{\bm{a}}^{T}-\hat{\bm{a}}\hat{\bm{a}}^{\dagger}\hat{\Lambda}$. 

\subsection{Second mixed product\label{subsec:Mixed-product-2}}

\begin{eqnarray}
 &  & \fl:\hat{\gamma}\Biggl\{\hat{\gamma}\hat{\Lambda}\Biggr\}:=\left(\begin{array}{cc}
\bm{T}_{1}^{'}-\left(\hat{\Lambda}\hat{\bm{a}}^{\dagger T}\hat{\bm{a}}^{T}\right)^{T}-\left(\hat{\bm{a}}\hat{\Lambda}\hat{\bm{a}}^{T}\right)^{T} & \rmi\left[\bm{T}_{2}^{'}-\hat{\bm{a}}^{\dagger T}\hat{\bm{a}}^{T}\hat{\Lambda}+\hat{\bm{a}}^{\dagger T}\hat{\Lambda}\hat{\bm{a}}^{\dagger}\right]\\
\rmi\left[\bm{T}_{1}^{'}+\left(\hat{\Lambda}\hat{\bm{a}}^{\dagger T}\hat{\bm{a}}^{T}\right)^{T}+\left(\hat{\bm{a}}\hat{\Lambda}\hat{\bm{a}}^{T}\right)^{T}\right] & \bm{T}_{2}^{'}+\hat{\bm{a}}^{\dagger T}\hat{\bm{a}}^{T}\hat{\Lambda}-\hat{\bm{a}}^{\dagger T}\hat{\Lambda}\hat{\bm{a}}^{\dagger}
\end{array}\right)\label{eq:ExpSMP}
\end{eqnarray}
where $\bm{T}_{1}^{'}=+\hat{\bm{a}}^{\dagger T}\hat{\bm{a}}^{T}\hat{\Lambda}+\hat{\bm{a}}^{\dagger T}\hat{\Lambda}\hat{\bm{a}}^{\dagger}$
and $\bm{T}_{2}^{'}=-\left(\hat{\Lambda}\hat{\bm{a}}^{\dagger T}\hat{\bm{a}}^{T}\right)^{T}+\left(\hat{\bm{a}}\hat{\Lambda}\hat{\bm{a}}^{T}\right)^{T}$.

\subsection{Normally ordered products \label{subsec:Normally-ordered-products}}

\begin{eqnarray}
 &  & \fl:\hat{\underline{\gamma}}\hat{\underline{\gamma}}^{T}\hat{\Lambda}:=\left(\begin{array}{cc}
\bm{T}_{3}+\hat{\bm{a}}^{\dagger T}\hat{\Lambda}\hat{\bm{a}}^{T}+\hat{\bm{a}}^{\dagger T}\hat{\bm{a}}^{\dagger}\hat{\Lambda} & -\rmi\left[\bm{T}_{4}+\hat{\bm{a}}^{\dagger T}\hat{\Lambda}\hat{\bm{a}}^{T}-\hat{\bm{a}}^{\dagger T}\hat{\bm{a}}^{\dagger}\hat{\Lambda}\right]\\
-\rmi\left[\bm{T}_{3}-\hat{\bm{a}}^{\dagger T}\hat{\Lambda}\hat{\bm{a}}^{T}-\hat{\bm{a}}^{\dagger T}\hat{\bm{a}}^{\dagger}\hat{\Lambda}\right] & -\left[\bm{T}_{4}-\hat{\bm{a}}^{\dagger T}\hat{\Lambda}\hat{\bm{a}}^{T}+\hat{\bm{a}}^{\dagger T}\hat{\bm{a}}^{\dagger}\hat{\Lambda}\right]
\end{array}\right),\label{eq:ENOP}
\end{eqnarray}
where $\bm{T}_{3}=\hat{\Lambda}\hat{\bm{a}}\hat{\bm{a}}^{T}-\left(\hat{\bm{a}}^{\dagger T}\hat{\Lambda}\hat{\bm{a}}^{T}\right)^{T}$
and $\bm{T}_{4}=\hat{\Lambda}\hat{\bm{a}}\hat{\bm{a}}^{T}+\left(\hat{\bm{a}}^{\dagger T}\hat{\Lambda}\hat{\bm{a}}^{T}\right)^{T}$.

\subsection{Anti-normally ordered products \label{subsec:Anti-normally-ordered-products}}

\begin{eqnarray}
 &  & \fl\left\{ \hat{\underline{\gamma}}\hat{\underline{\gamma}}^{T}\hat{\Lambda}\right\} =\left(\begin{array}{cc}
\bm{T}_{5}-\left(\hat{\bm{a}}\hat{\Lambda}\hat{\bm{a}}^{\dagger}\right)^{T}+\hat{\Lambda}\hat{\bm{a}}^{\dagger T}\hat{\bm{a}}^{\dagger} & -\rmi\left[\bm{T}_{6}-\left(\hat{\bm{a}}\hat{\Lambda}\hat{\bm{a}}^{\dagger}\right)^{T}-\hat{\Lambda}\hat{\bm{a}}^{\dagger T}\hat{\bm{a}}^{\dagger}\right]\\
-\rmi\left[\bm{T}_{5}+\left(\hat{\bm{a}}\hat{\Lambda}\hat{\bm{a}}^{\dagger}\right)^{T}-\hat{\Lambda}\hat{\bm{a}}^{\dagger T}\hat{\bm{a}}^{\dagger}\right] & -\left[\bm{T}_{6}+\left(\hat{\bm{a}}\hat{\Lambda}\hat{\bm{a}}^{\dagger}\right)^{T}+\hat{\Lambda}\hat{\bm{a}}^{\dagger T}\hat{\bm{a}}^{\dagger}\right]
\end{array}\right),\label{eq:ExpANOP}
\end{eqnarray}
where $\bm{T}_{5}=\hat{\bm{a}}\hat{\bm{a}}^{T}\hat{\Lambda}+\hat{\bm{a}}\hat{\Lambda}\hat{\bm{a}}^{\dagger}$
and $\bm{T}_{6}=\hat{\bm{a}}\hat{\bm{a}}^{T}\hat{\Lambda}-\hat{\bm{a}}\hat{\Lambda}\hat{\bm{a}}^{\dagger}$.

\subsection{Unordered products \label{subsec:UnOrderedProduct}}

\begin{eqnarray}
 &  & \fl\hat{\underline{\gamma}}\hat{\underline{\gamma}}^{T}\hat{\Lambda}=\left[\begin{array}{cc}
\bm{T}_{7}+\hat{\bm{a}}^{\dagger T}\hat{\bm{a}}^{\dagger}\hat{\Lambda}+\hat{\bm{a}}^{\dagger}{}^{T}\hat{\bm{a}}^{T}\hat{\Lambda} & \rmi\left[\bm{T}_{8}+\hat{\bm{a}}^{\dagger T}\hat{\bm{a}}^{\dagger}\hat{\Lambda}-\hat{\bm{a}}^{\dagger}{}^{T}\hat{\bm{a}}^{T}\hat{\Lambda}\right]\\
\rmi\left[-\bm{T}_{7}+\hat{\bm{a}}^{\dagger T}\hat{\bm{a}}^{\dagger}\hat{\Lambda}+\hat{\bm{a}}^{\dagger}{}^{T}\hat{\bm{a}}^{T}\hat{\Lambda}\right] & \bm{T}_{8}-\hat{\bm{a}}^{\dagger T}\hat{\bm{a}}^{\dagger}\hat{\Lambda}+\hat{\bm{a}}^{\dagger}{}^{T}\hat{\bm{a}}^{T}\hat{\Lambda}
\end{array}\right]\label{eq:ExpUn-Or}
\end{eqnarray}
where $\bm{T}_{7}=\hat{\bm{a}}\hat{\bm{a}}^{\dagger}\hat{\Lambda}+\hat{\bm{a}}\hat{\bm{a}}^{T}\hat{\Lambda}$
and $\bm{T}_{8}=\hat{\bm{a}}\hat{\bm{a}}^{\dagger}\hat{\Lambda}-\hat{\bm{a}}\hat{\bm{a}}^{T}\hat{\Lambda}$.

\section{Majorana differential identities for un-normalized ordered products\label{sec:AppendixMajoranaDiffId}}

In this Section we give the detailed proofs of the un-normalized,
ordered product differential identities shown in Section \ref{sec:Majorana-differential-properties}.
We first give the proof for the un-normalized Majorana differential
identities and then in the next section we use these identities in
order to obtain the normalized ones. 

The following differential identities are given considering the different
types of ordering of the Majorana variables and Gaussian operators.
The explicit form of these are given in \ref{sec:AppendixFormOrdering}.
These identities uses the un-normalized Gaussian operators given in
(\ref{eq:GOY}). 

\subsection{Mixed products}

Here we are considering products of the Majorana variables and Gaussian
operators of the form $\Biggl\{\hat{\underline{\gamma}}:\hat{\underline{\gamma}}^{T}\hat{\Lambda}^{\left(u\right)}:\Biggr\}$.
In this case the differential identity is:

\[
\Biggl\{\hat{\underline{\gamma}}:\hat{\underline{\gamma}}^{T}\hat{\Lambda}^{\left(u\right)}:\Biggr\}=\rmi\left[-\underline{\underline{{\cal I}}}\left(2\underline{\underline{Y}}+\underline{\underline{\mathcal{I}}}\right)\frac{\rmd}{\rmd\underline{\underline{Y}}}+2\underline{\underline{{\cal I}}}\right]\hat{\Lambda}^{\left(u\right)}.
\]

\textbf{\emph{Proof.}} To prove this identity we proceed by using
Grassmann variables and fermionic coherent states. We also use the
following result \cite{Corney_PD_JPA_2006_GR_fermions} that is already
known, but with more details included for clarity and completeness:
\begin{eqnarray}
\fl\Biggl\{\underline{\widehat{a}}:\widehat{\underline{a}}^{\dagger}\hat{\Lambda}^{\left(u\right)}:\Biggr\} & = & \int\rmd\underline{\gamma}\rmd\underline{\beta}\rmd\underline{\alpha}\rmd\underline{\epsilon}\Biggl|\bgamma\Biggr\rangle\Biggl\langle\bgamma\Biggr|\Biggl\{\underline{\widehat{a}}:\Biggl|\bbeta\Biggr\rangle\Biggl\langle\bbeta\Biggr|\widehat{\underline{a}}^{\dagger}\widehat{\Lambda}^{\left(u\right)}\Biggl|\balpha\Biggr\rangle\Biggl\langle\balpha\Biggr|:\Biggr\}\Biggl|\bepsilon\Biggr\rangle\Biggl\langle\bepsilon\Biggr|\nonumber \\
 & = & \int\rmd\underline{\gamma}\rmd\underline{\beta}\rmd\underline{\alpha}\rmd\underline{\epsilon}\Biggl|\bgamma\Biggr\rangle\Biggl\langle\bepsilon\Biggr|\exp\left[-\bar{\balpha}\balpha-\bar{\bbeta}\bbeta-\frac{1}{2}\bar{\bgamma}\bgamma-\frac{1}{2}\bar{\bepsilon}\bepsilon+\bar{\bgamma}\bbeta+\bar{\balpha}\bepsilon\right]\nonumber \\
 & \times & \left[\begin{array}{c}
\bbeta\\
\bar{\balpha}
\end{array}\right]\left[\begin{array}{cc}
\bar{\bbeta} & \balpha\end{array}\right]\exp\left[-\frac{1}{2}\left[\begin{array}{cc}
\bar{\bbeta} & \balpha\end{array}\right]\left(\underline{\underline{\mu}}+\underline{\underline{J}}\right)\left[\begin{array}{c}
\balpha\\
\bar{\bbeta}
\end{array}\right]\right]\nonumber \\
 & = & \int\rmd\underline{\gamma}\rmd\underline{\beta}\rmd\underline{\alpha}\rmd\underline{\epsilon}\Biggl|\bgamma\Biggr\rangle\Biggl\langle\bepsilon\Biggr|\exp\left[-\bar{\balpha}\balpha-\bar{\bbeta}\bbeta-\frac{1}{2}\bar{\bgamma}\bgamma-\frac{1}{2}\bar{\bepsilon}\bepsilon+\bar{\bgamma}\bbeta+\bar{\balpha}\bepsilon\right]\nonumber \\
 & \times & \left[\underline{\underline{J}}\left(\underline{\underline{\mu}}+\underline{\underline{J}}\right)\left[\begin{array}{c}
\balpha\\
\bar{\bbeta}
\end{array}\right]\left[\begin{array}{cc}
\bar{\bbeta} & \balpha\end{array}\right]-\underline{\underline{J}}\right]\exp\left[-\left[\begin{array}{cc}
\bar{\bbeta} & \balpha\end{array}\right]\frac{\left(\underline{\underline{\mu}}+\underline{\underline{J}}\right)}{2}\left[\begin{array}{c}
\balpha\\
\bar{\bbeta}
\end{array}\right]\right].\nonumber \\
\label{eq:Eq.B42}
\end{eqnarray}
Here $\left|\balpha\right\rangle $, $\left|\bbeta\right\rangle $,
$\left|\bgamma\right\rangle $ and $\left|\bepsilon\right\rangle $
denote fermionic coherent states, while $\balpha$, $\bbeta$, $\bgamma$
and $\bepsilon$ are the corresponding Grassmann variables, and $\rmd\underline{\gamma}$,
$\rmd\underline{\beta},$ $\rmd\underline{\alpha}$ and $\rmd\underline{\epsilon}$
are the corresponding $2M$ integration measures. In order to obtain
the identity given in (\ref{eq:Eq.B42}) we have used an integration
by parts and the properties of Grassmann calculus, as well as the
eigenvalue properties for the fermionic coherent states in the form
$\hat{\bm{a}}\left|\balpha\right\rangle =\balpha\left|\balpha\right\rangle $.
We have also used the resolution of unity or completeness identity
of the fermionic coherent states, together with the inner product
property of coherent states which are given below:
\begin{eqnarray}
\int d\underline{\alpha}\left|\balpha\right\rangle \left\langle \balpha\right| & = & \mathbf{I},\label{eq:ResUnitCS}\\
\left\langle \bbeta\vert\balpha\right\rangle  & = & \exp\left[\bar{\bbeta}\balpha-\left(\bar{\bbeta}\bbeta+\bar{\balpha}\balpha\right)/2\right].\label{eq:InnerProdCS}
\end{eqnarray}
One important step is to express the variables $\left[\begin{array}{c}
\bbeta\\
\bar{\balpha}
\end{array}\right]$ in the form $\left[\begin{array}{c}
\balpha\\
\bar{\bbeta}
\end{array}\right],$ which is the one given in the exponential of the Gaussian operators.
This is done by using the following identities, which use the derivative
of the Gaussian operator and Grassmann calculus:
\[
\left[\begin{array}{c}
\bbeta\\
\bar{\balpha}
\end{array}\right]\exp\left[-\bar{\balpha}\balpha-\bar{\bbeta}\bbeta\right]=\left[\begin{array}{c}
-\frac{\partial}{\partial\overline{\bbeta}}\\
\frac{\partial}{\partial\balpha}
\end{array}\right]\exp\left[-\bar{\balpha}\balpha-\bar{\bbeta}\bbeta\right].
\]
\begin{eqnarray*}
 &  & \fl\exp\left[-\bar{\balpha}\balpha-\bar{\bbeta}\bbeta\right]\left[\begin{array}{c}
\frac{\partial}{\partial\overline{\bbeta}}\\
-\frac{\partial}{\partial\balpha}
\end{array}\right]\exp\left[-\frac{1}{2}\left[\begin{array}{cc}
\bar{\bbeta} & \balpha\end{array}\right]\left(\underline{\underline{\mu}}+\underline{\underline{J}}\right)\left[\begin{array}{c}
\balpha\\
\bar{\bbeta}
\end{array}\right]\right]\left[\begin{array}{cc}
\bar{\bbeta} & \balpha\end{array}\right]\\
 &  & \fl=\exp\left[-\bar{\balpha}\balpha-\bar{\bbeta}\bbeta\right]\left[\underline{\underline{J}}\left(\underline{\underline{\mu}}+\underline{\underline{J}}\right)\left[\begin{array}{c}
\balpha\\
\bar{\bbeta}
\end{array}\right]\left[\begin{array}{cc}
\bar{\bbeta} & \balpha\end{array}\right]-\underline{\underline{J}}\right]\\
 &  & \fl\times\exp\left[-\frac{1}{2}\left[\begin{array}{cc}
\bar{\bbeta} & \balpha\end{array}\right]\left(\underline{\underline{\mu}}+\underline{\underline{J}}\right)\left[\begin{array}{c}
\balpha\\
\bar{\bbeta}
\end{array}\right]\right].
\end{eqnarray*}
Hence, using integration by parts and the above identities, (\ref{eq:Eq.B42})
is obtained. 

Next, we wish to express the above identities in terms of Majorana
operators. Thus without changing the order of the ladder operators,
we use the identities that relate the Majorana operators with the
ladder operators given in (\ref{eq:OpMajUn}), as well as the expression
given in (\ref{eq:YMat}). We also multiply the right side of both
sides of equation (\ref{eq:Eq.B42}) by $2\underline{\underline{U_{0}}}^{-1}$,
 obtaining:

\begin{eqnarray}
\fl\Biggl\{\underline{\underline{U_{0}}}^{-1}\hat{\underline{\gamma}}:\hat{\underline{\gamma}}^{T}\hat{\Lambda}^{\left(u\right)}:\Biggr\} & = & \int\rmd\underline{\gamma}\rmd\underline{\beta}\rmd\underline{\alpha}\rmd\underline{\epsilon}\Biggl|\bgamma\Biggr\rangle\Biggl\langle\bepsilon\Biggr|\exp\left[-\bar{\balpha}\balpha-\bar{\bbeta}\bbeta-\frac{1}{2}\bar{\bgamma}\bgamma-\frac{1}{2}\bar{\bepsilon}\bepsilon+\bar{\bgamma}\bbeta+\bar{\balpha}\bepsilon\right]\nonumber \\
 &  & \times\left[-\rmi2\underline{\underline{J}}\underline{\underline{U_{0}}}^{-1}\left(2\underline{\underline{Y}}+\underline{\underline{\mathcal{I}}}\right)\underline{\underline{U_{0}}}\left[\begin{array}{c}
\balpha\\
\bar{\bbeta}
\end{array}\right]\left[\begin{array}{cc}
\bar{\bbeta} & \balpha\end{array}\right]\underline{\underline{U_{0}}}^{-1}-2\underline{\underline{J}}\underline{\underline{U_{0}}}^{-1}\right]\nonumber \\
 &  & \times\exp\left[\frac{\rmi}{2}\left[\begin{array}{cc}
\bar{\bbeta} & \balpha\end{array}\right]\underline{\underline{U_{0}}}^{-1}\left(2\underline{\underline{Y}}+\underline{\underline{\mathcal{I}}}\right)\underline{\underline{U_{0}}}\left[\begin{array}{c}
\balpha\\
\bar{\bbeta}
\end{array}\right]\right].\label{eq:MajOpT}
\end{eqnarray}
We notice that the second and third lines of (\ref{eq:MajOpT}) can
be written in terms of the derivative of the un-normalized Gaussian
operator itself with respect to $\underline{\underline{Y}}$. Hence
we get the following result:
\begin{eqnarray}
\fl\Biggl\{\underline{\underline{U_{0}}}^{-1}\hat{\underline{\gamma}}:\hat{\underline{\gamma}}^{T}\hat{\Lambda}^{\left(u\right)}:\Biggr\} & = & \int\rmd\underline{\gamma}\rmd\underline{\beta}\rmd\underline{\alpha}\rmd\underline{\epsilon}\Biggl|\bgamma\Biggr\rangle\Biggl\langle\bepsilon\Biggr|\exp\left[-\bar{\balpha}\balpha-\bar{\bbeta}\bbeta-\frac{1}{2}\bar{\bgamma}\bgamma-\frac{1}{2}\bar{\bepsilon}\bepsilon+\bar{\bgamma}\bbeta+\bar{\balpha}\bepsilon\right]\nonumber \\
 & \times & \left[\overline{\underline{\underline{I}}}\underline{\underline{U_{0}}}^{-1}\left(2\underline{\underline{Y}}+\underline{\underline{\mathcal{I}}}\right)\frac{\rmd}{\rmd\underline{\underline{Y}}}-2\underline{\underline{\overline{I}}}\underline{\underline{U_{0}}}^{-1}\right]\nonumber \\
 & \times & \exp\left[\frac{\rmi}{2}\left[\begin{array}{cc}
\bar{\bbeta} & \balpha\end{array}\right]\underline{\underline{U_{0}}}^{-1}\left(2\underline{\underline{Y}}+\underline{\underline{\mathcal{I}}}\right)\underline{\underline{U_{0}}}\left[\begin{array}{c}
\balpha\\
\bar{\bbeta}
\end{array}\right]\right]\\
 & = & N\left[\overline{\underline{\underline{I}}}\underline{\underline{U_{0}}}^{-1}\left(2\underline{\underline{Y}}+\underline{\underline{\mathcal{I}}}\right)\frac{\rmd}{\rmd\underline{\underline{Y}}}-2\underline{\underline{\overline{I}}}\underline{\underline{U_{0}}}^{-1}\right]\hat{\Lambda}^{\left(u\right)}\left(\underline{\underline{Y}}\right).\label{eq:MajIdMPCS}
\end{eqnarray}
In order to obtain the last line of (\ref{eq:MajIdMPCS}) we have
used the resolution of unity of coherent states given in (\ref{eq:ResUnitCS})
and the inner product of fermionic coherent states given in (\ref{eq:InnerProdCS})
in order to express the Majorana Gaussian operator as:
\begin{eqnarray}
\fl\hat{\Lambda}^{\left(u\right)}\left(\underline{\underline{Y}}\right) & = & \int\rmd\underline{\gamma}\rmd\underline{\beta}\rmd\underline{\alpha}\rmd\underline{\epsilon}\Biggl|\bgamma\Biggr\rangle\Biggl\langle\bepsilon\Biggr|\exp\left[-\bar{\balpha}\balpha-\bar{\bbeta}\bbeta-\frac{1}{2}\bar{\bgamma}\bgamma-\frac{1}{2}\bar{\bepsilon}\bepsilon+\bar{\bgamma}\bbeta+\bar{\balpha}\bepsilon\right]\nonumber \\
 & \times & \exp\left[\frac{\rmi}{2}\left[\begin{array}{cc}
\bar{\bbeta} & \balpha\end{array}\right]\underline{\underline{U_{0}}}^{-1}\left(2\underline{\underline{Y}}+\underline{\underline{\mathcal{I}}}\right)\underline{\underline{U_{0}}}\left[\begin{array}{c}
\balpha\\
\bar{\bbeta}
\end{array}\right]\right].\label{eq:MajGO_ResUnit}
\end{eqnarray}
 On multiplying on the left both sides of (\ref{eq:MajIdMPCS}), by
$\underline{\underline{U_{0}}}$, we get:
\begin{eqnarray}
\Biggl\{\hat{\underline{\gamma}}:\hat{\underline{\gamma}}^{T}\hat{\Lambda}^{\left(u\right)}:\Biggr\} & = & \left[\underline{\underline{U_{0}}}\overline{\underline{\underline{I}}}\underline{\underline{U_{0}}}^{-1}\left(2\underline{\underline{Y}}+\underline{\underline{\mathcal{I}}}\right)\frac{\rmd}{\rmd\underline{\underline{Y}}}-2\underline{\underline{U_{0}}}\underline{\underline{\overline{I}}}\underline{\underline{U_{0}}}^{-1}\right]\hat{\Lambda}^{\left(u\right)}\nonumber \\
 & = & \rmi\left[-\underline{\underline{{\cal I}}}\left(2\underline{\underline{Y}}+\underline{\underline{\mathcal{I}}}\right)\frac{\rmd}{\rmd\underline{\underline{Y}}}+2\underline{\underline{{\cal I}}}\right]\hat{\Lambda}^{\left(u\right)}.\label{eq:DifIY}
\end{eqnarray}
Here we have used that from the definition of $\underline{\underline{{\cal I}}}$
given in (\ref{eq:UnitFI}) we get $-i\underline{\underline{{\cal I}}}=\underline{\underline{U_{0}}}\overline{\underline{\underline{I}}}\underline{\underline{U_{0}}}^{-1}$.
Therefore we have proved the differential identity given in (\ref{eq:MajDiffIdY}).

\subsection{Normally ordered products }

\[
:\hat{\underline{\gamma}}\hat{\underline{\gamma}}^{T}\hat{\Lambda}^{\left(u\right)}:=\rmi\frac{d}{d\underline{\underline{Y}}}\hat{\Lambda}^{\left(u\right)}.
\]
\textbf{\emph{Proof}}. Analogous to the previous case we make use
of the eigenvalue property of the fermionic coherent states as well
as the completeness identity and the inner product of coherent states.
Thus, we get:
\begin{eqnarray*}
\fl:\underline{\widehat{a}}\widehat{\underline{a}}^{\dagger}\hat{\Lambda}^{\left(u\right)}: & = & \int\rmd\underline{\gamma}\rmd\underline{\beta}\rmd\underline{\alpha}\rmd\underline{\epsilon}\Biggl|\bgamma\Biggr\rangle\Biggl\langle\bepsilon\Biggr|\exp\left[-\bar{\balpha}\balpha-\bar{\bbeta}\bbeta-\frac{1}{2}\bar{\bgamma}\bgamma-\frac{1}{2}\bar{\bepsilon}\bepsilon+\bar{\bgamma}\bbeta+\bar{\balpha}\bepsilon\right]\\
 &  & \times\left[\begin{array}{c}
\balpha\\
\bar{\bbeta}
\end{array}\right]\left[\begin{array}{cc}
\bar{\bbeta} & \balpha\end{array}\right]\exp\left[-\frac{1}{2}\left[\begin{array}{cc}
\bar{\bbeta} & \balpha\end{array}\right]\left(\underline{\underline{\mu}}+\underline{\underline{J}}\right)\left[\begin{array}{c}
\balpha\\
\bar{\bbeta}
\end{array}\right]\right].
\end{eqnarray*}

Next, without changing the order of the Grassmann variables, we can
write the above expression in terms of Majorana operators using equations
(\ref{eq:OpMajUn}) and (\ref{eq:YMat}). We also multiply both sides
of the above equation on the left by $2\underline{\underline{U_{0}}}$
and on the right by $\underline{\underline{U_{0}}}^{-1}$, thus obtaining:
\begin{eqnarray*}
\fl:\hat{\underline{\gamma}}\hat{\underline{\gamma}}^{T}\hat{\Lambda}^{\left(u\right)}: & = & \int\rmd\underline{\gamma}\rmd\underline{\beta}\rmd\underline{\alpha}\rmd\underline{\epsilon}\Biggl|\bgamma\Biggr\rangle\Biggl\langle\bepsilon\Biggr|\exp\left[-\bar{\balpha}\balpha-\bar{\bbeta}\bbeta-\frac{1}{2}\bar{\bgamma}\bgamma-\frac{1}{2}\bar{\bepsilon}\bepsilon+\bar{\bgamma}\bbeta+\bar{\balpha}\bepsilon\right]\\
 &  & \times2\underline{\underline{U_{0}}}\left[\begin{array}{c}
\balpha\\
\bar{\bbeta}
\end{array}\right]\left[\begin{array}{cc}
\bar{\bbeta} & \balpha\end{array}\right]\underline{\underline{U_{0}}}^{-1}\exp\left[\frac{\rmi}{2}\left[\begin{array}{cc}
\bar{\bbeta} & \balpha\end{array}\right]\underline{\underline{U_{0}}}^{-1}\left(2\underline{\underline{Y}}+\underline{\underline{\mathcal{I}}}\right)\underline{\underline{U_{0}}}\left[\begin{array}{c}
\balpha\\
\bar{\bbeta}
\end{array}\right]\right]\\
 & = & \int\rmd\underline{\gamma}\rmd\underline{\beta}\rmd\underline{\alpha}\rmd\underline{\epsilon}\Biggl|\bgamma\Biggr\rangle\Biggl\langle\bepsilon\Biggr|\exp\left[-\bar{\balpha}\balpha-\bar{\bbeta}\bbeta-\frac{1}{2}\bar{\bgamma}\bgamma-\frac{1}{2}\bar{\bepsilon}\bepsilon+\bar{\bgamma}\bbeta+\bar{\balpha}\bepsilon\right]\\
 &  & \times\left(\frac{d}{d\underline{\underline{Y}}}\exp\left[\frac{\rmi}{2}\left[\begin{array}{cc}
\bar{\bbeta} & \balpha\end{array}\right]\underline{\underline{U_{0}}}^{-1}\left(2\underline{\underline{Y}}+\underline{\underline{\mathcal{I}}}\right)\underline{\underline{U_{0}}}\left[\begin{array}{c}
\balpha\\
\bar{\bbeta}
\end{array}\right]\right]\right).\\
 & = & \rmi\frac{d}{d\underline{\underline{Y}}}\hat{\Lambda}^{\left(u\right)}.
\end{eqnarray*}
Here we have expressed the results of the first two lines as the derivative
of the Gaussian operator an we have used the expression of the Majorana
Gaussian operator given in (\ref{eq:MajGO_ResUnit}). Therefore we
have proved the differential identity given in (\ref{eq:UnNormNormalDifId}).

\subsection{Anti-normally ordered products }

\[
\left\{ \underline{\widehat{\gamma}}\widehat{\underline{\gamma}}^{T}\hat{\Lambda}^{\left(u\right)}\right\} =\rmi\underline{\underline{\mathcal{I}}}\left\{ \left(2\underline{\underline{Y}}+\underline{\underline{\mathcal{I}}}\right)\frac{\rmd}{\rmd\underline{\underline{Y}}}\hat{\Lambda}^{u}-2\hat{\Lambda}^{u}\right\} \left(2\underline{\underline{Y}}+\underline{\underline{\mathcal{I}}}\right)\underline{\underline{\mathcal{I}}}.
\]

\textbf{\emph{Proof.}} As in the previous cases we use the eigenvalue
property of the fermionic coherent states as well as the completeness
identity and the inner product of coherent states in order to write
$\left\{ \underline{\widehat{a}}\widehat{\underline{a}}^{\dagger}\hat{\Lambda}^{\left(u\right)}\right\} $
in terms of Grassmann variables, obtaining:
\begin{eqnarray}
\fl\left\{ \underline{\widehat{a}}\widehat{\underline{a}}^{\dagger}\hat{\Lambda}^{\left(u\right)}\right\}  & = & \int\rmd\underline{\gamma}\rmd\underline{\beta}\rmd\underline{\alpha}\rmd\underline{\epsilon}\Biggl|\bgamma\Biggr\rangle\Biggl\langle\bgamma\Biggr|\Biggl\{\underline{\widehat{a}}\widehat{\underline{a}}^{\dagger}\Biggl|\bbeta\Biggr\rangle\Biggl\langle\bbeta\Biggr|\widehat{\Lambda}^{\left(u\right)}\Biggl|\balpha\Biggr\rangle\Biggl\langle\balpha\Biggr|:\Biggr\}\Biggl|\bepsilon\Biggr\rangle\Biggl\langle\bepsilon\Biggr|\nonumber \\
 & = & \int\rmd\underline{\gamma}\rmd\underline{\beta}\rmd\underline{\alpha}\rmd\underline{\epsilon}\Biggl|\bgamma\Biggr\rangle\Biggl\langle\bepsilon\Biggr|\exp\left[-\bar{\balpha}\balpha-\bar{\bbeta}\bbeta-\frac{1}{2}\bar{\bgamma}\bgamma-\frac{1}{2}\bar{\bepsilon}\bepsilon+\bar{\bgamma}\bbeta+\bar{\balpha}\bepsilon\right]\nonumber \\
 &  & \times\exp\left[-\frac{1}{2}\left[\begin{array}{cc}
\bar{\bbeta} & \balpha\end{array}\right]\left(\underline{\underline{\mu}}+\underline{\underline{J}}\right)\left[\begin{array}{c}
\balpha\\
\bar{\bbeta}
\end{array}\right]\right]\left[\begin{array}{c}
\bbeta\\
\bar{\balpha}
\end{array}\right]\left[\begin{array}{cc}
\bar{\balpha} & \bbeta\end{array}\right].\label{eq:AntiNormalDPGv}
\end{eqnarray}
Next we wish to express the variables $\left[\begin{array}{c}
\bbeta\\
\bar{\balpha}
\end{array}\right]$ and $\left[\begin{array}{cc}
\bar{\balpha} & \bbeta\end{array}\right]$ in the form given in the exponential of the Gaussian operator. In
this case we will also use integration by parts but it will be performed
twice. To this end we will use the following identities, which make
use of Grassmann calculus:
\begin{eqnarray*}
 &  & \fl\exp\left[-\frac{1}{2}\left[\begin{array}{cc}
\bar{\bbeta} & \balpha\end{array}\right]\left(\underline{\underline{\mu}}+\underline{\underline{J}}\right)\left[\begin{array}{c}
\balpha\\
\bar{\bbeta}
\end{array}\right]\right]\left[\begin{array}{c}
\bbeta\\
\bar{\balpha}
\end{array}\right]\left[\begin{array}{cc}
\bar{\balpha} & \bbeta\end{array}\right]\exp\left[-\bar{\balpha}\balpha-\bar{\bbeta}\bbeta\right]=\\
 &  & \fl\exp\left[-\frac{1}{2}\left[\begin{array}{cc}
\bar{\bbeta} & \balpha\end{array}\right]\left(\underline{\underline{\mu}}+\underline{\underline{J}}\right)\left[\begin{array}{c}
\balpha\\
\bar{\bbeta}
\end{array}\right]\right]\left[\begin{array}{c}
-\frac{\partial}{\partial\overline{\bbeta}}\\
\frac{\partial}{\partial\balpha}
\end{array}\right]\left[\begin{array}{cc}
\frac{\partial}{\partial\balpha} & -\frac{\partial}{\partial\overline{\bbeta}}\end{array}\right]\exp\left[-\bar{\balpha}\balpha-\bar{\bbeta}\bbeta\right]
\end{eqnarray*}
\begin{eqnarray*}
 &  & \fl\exp\left[-\bar{\balpha}\balpha-\bar{\bbeta}\bbeta\right]\left[\begin{array}{c}
-\frac{\partial}{\partial\overline{\bbeta}}\\
\frac{\partial}{\partial\balpha}
\end{array}\right]\left[\begin{array}{cc}
\frac{\partial}{\partial\balpha} & -\frac{\partial}{\partial\overline{\bbeta}}\end{array}\right]\exp\left[-\frac{1}{2}\left[\begin{array}{cc}
\bar{\bbeta} & \balpha\end{array}\right]\left(\underline{\underline{\mu}}+\underline{\underline{J}}\right)\left[\begin{array}{c}
\balpha\\
\bar{\bbeta}
\end{array}\right]\right]\\
 &  & \fl=\exp\left[-\bar{\balpha}\balpha-\bar{\bbeta}\bbeta\right]\left[\underline{\underline{J}}\left(\underline{\underline{\mu}}+\underline{\underline{J}}\right)\left[\begin{array}{c}
\balpha\\
\bar{\bbeta}
\end{array}\right]\left[\begin{array}{cc}
\bar{\bbeta} & \balpha\end{array}\right]-\underline{\underline{J}}\right]\left(\underline{\underline{\mu}}+\underline{\underline{J}}\right)\underline{\underline{J}}\\
 &  & \fl\times\exp\left[-\frac{1}{2}\left[\begin{array}{cc}
\bar{\bbeta} & \balpha\end{array}\right]\left(\underline{\underline{\mu}}+\underline{\underline{J}}\right)\left[\begin{array}{c}
\balpha\\
\bar{\bbeta}
\end{array}\right]\right].
\end{eqnarray*}
The above expressions allow us to perform an integration by parts
twice in (\ref{eq:AntiNormalDPGv}) obtaining:
\begin{eqnarray*}
\fl\left\{ \underline{\widehat{a}}\widehat{\underline{a}}^{\dagger}\hat{\Lambda}^{\left(u\right)}\right\}  & = & \int\rmd\underline{\gamma}\rmd\underline{\beta}\rmd\underline{\alpha}\rmd\underline{\epsilon}\Biggl|\bgamma\Biggr\rangle\Biggl\langle\bepsilon\Biggr|\exp\left[-\bar{\balpha}\balpha-\bar{\bbeta}\bbeta-\frac{1}{2}\bar{\bgamma}\bgamma-\frac{1}{2}\bar{\bepsilon}\bepsilon+\bar{\bgamma}\bbeta+\bar{\balpha}\bepsilon\right]\\
 &  & \times\left[\underline{\underline{J}}\left(\underline{\underline{\mu}}+\underline{\underline{J}}\right)\left[\begin{array}{c}
\balpha\\
\bar{\bbeta}
\end{array}\right]\left[\begin{array}{cc}
\bar{\bbeta} & \balpha\end{array}\right]-\underline{\underline{J}}\right]\left(\underline{\underline{\mu}}+\underline{\underline{J}}\right)\underline{\underline{J}}\\
 &  & \times\exp\left[-\frac{1}{2}\left[\begin{array}{cc}
\bar{\bbeta} & \balpha\end{array}\right]\left(\underline{\underline{\mu}}+\underline{\underline{J}}\right)\left[\begin{array}{c}
\balpha\\
\bar{\bbeta}
\end{array}\right]\right].
\end{eqnarray*}

Next since we wish to express the identity in terms of Majorana operators
and an anti-symmetric matrix, without changing the order of the ladder
operators, we use the identities that relate the ladder operators
with the Majorana operators given in (\ref{eq:OpMajUn}), as well
the relation given in (\ref{eq:YMat}) and (\ref{eq:UnitFI}). We
also multiply both sides of the above equation on the left by $2\underline{\underline{U_{0}}}$
and on the right by $\underline{\underline{U_{0}}}^{-1}$, thus obtaining:
\begin{eqnarray*}
\fl\left\{ \underline{\widehat{\gamma}}\widehat{\underline{\gamma}}^{T}\hat{\Lambda}^{\left(u\right)}\right\}  & = & \int\rmd\underline{\gamma}\rmd\underline{\beta}\rmd\underline{\alpha}\rmd\underline{\epsilon}\Biggl|\bgamma\Biggr\rangle\Biggl\langle\bepsilon\Biggr|\exp\left[-\bar{\balpha}\balpha-\bar{\bbeta}\bbeta-\frac{1}{2}\bar{\bgamma}\bgamma-\frac{1}{2}\bar{\bepsilon}\bepsilon+\bar{\bgamma}\bbeta+\bar{\balpha}\bepsilon\right]\\
 &  & \times2\left[\underline{\underline{\mathcal{I}}}\left(2\underline{\underline{Y}}+\underline{\underline{\mathcal{I}}}\right)\underline{\underline{U_{0}}}\left[\begin{array}{c}
\balpha\\
\bar{\bbeta}
\end{array}\right]\left[\begin{array}{cc}
\bar{\bbeta} & \balpha\end{array}\right]\underline{\underline{U_{0}}}^{-1}-\rmi\underline{\underline{\mathcal{I}}}\right]\left(2\underline{\underline{Y}}+\underline{\underline{\mathcal{I}}}\right)\underline{\underline{\mathcal{I}}}\\
 &  & \times\exp\left[\frac{\rmi}{2}\left[\begin{array}{cc}
\bar{\bbeta} & \balpha\end{array}\right]\underline{\underline{U_{0}}}^{-1}\left(2\underline{\underline{Y}}+\underline{\underline{\mathcal{I}}}\right)\underline{\underline{U_{0}}}\left[\begin{array}{c}
\balpha\\
\bar{\bbeta}
\end{array}\right]\right].
\end{eqnarray*}
 As in the previous cases we notice that the second and third line
of the above expressions can be expressed in terms of the derivative
of the un-normalized Gaussian operator with respect to $\underline{\underline{Y}}$.
We also use the expression for the un-normalized Gaussian operators
given in (\ref{eq:MajGO_ResUnit}). Therefore we get:
\begin{eqnarray*}
\left\{ \underline{\widehat{\gamma}}\widehat{\underline{\gamma}}^{T}\hat{\Lambda}^{\left(u\right)}\right\}  & = & \rmi\underline{\underline{\mathcal{I}}}\left\{ \left(2\underline{\underline{Y}}+\underline{\underline{\mathcal{I}}}\right)\frac{\rmd}{\rmd\underline{\underline{Y}}}\hat{\Lambda}^{u}-2\hat{\Lambda}^{u}\right\} \left(2\underline{\underline{Y}}+\underline{\underline{\mathcal{I}}}\right)\underline{\underline{\mathcal{I}}}.
\end{eqnarray*}
Thus we have proved the differential identity given in (\ref{eq:AntiNormDifIdUnGO}).

\section{Normalized Majorana differential identities\label{sec:AppendixDiffIdNormGO}}

Here we consider the normalized Gaussian operators given in (\ref{eq:NormGOX}).
We also used the change of variables given in (\ref{eq:ChangeVarYX}).

\subsection{First mixed product}

\[
\Biggl\{\hat{\underline{\gamma}}:\hat{\underline{\gamma}}^{T}\hat{\Lambda}:\Biggr\}=\rmi\left(\underline{\underline{X}}+\underline{\underline{{\cal I}}}\right)\frac{\rmd\hat{\Lambda}\left(\underline{\underline{X}}\right)}{\rmd\underline{\underline{X}}}\left(\underline{\underline{X}}-\underline{\underline{\mathcal{I}}}\right)-\rmi\left(\underline{\underline{X}}-\underline{\underline{{\cal I}}}\right)\hat{\Lambda}\left(\underline{\underline{X}}\right).
\]
\textbf{\emph{Proof.}} We make a change of variables using (\ref{eq:ChangeVarYX})
on the differential identities given in (\ref{eq:MajDiffIdY}) obtaining:
\begin{equation}
\fl\Biggl\{\hat{\underline{\gamma}}:\hat{\underline{\gamma}}^{T}\hat{\Lambda}:\Biggr\}=\rmi N\left(\underline{\underline{X}}\right)\left[-\underline{\underline{{\cal I}}}\left(2\left(\underline{\underline{X}}-\underline{\underline{\mathcal{I}}}\right)^{-1}-\underline{\underline{\mathcal{I}}}\right)\frac{\rmd}{\rmd\underline{\underline{Y}}}+2\underline{\underline{{\cal I}}}\right]\hat{\Lambda}^{\left(u\right)}\left(\underline{\underline{X}}\right).\label{eq:DifIndYX}
\end{equation}
Next, we use the chain rule in order to change the derivative, which
is:
\begin{equation}
\fl\frac{\rmd\hat{\Lambda}^{\left(u\right)}}{\rmd\underline{\underline{Y}}}=\frac{\rmd\hat{\Lambda}^{\left(u\right)}}{\rmd\underline{\underline{X}}}\frac{\rmd\underline{\underline{X}}}{\rmd\underline{\underline{Y}}}=-\left(\underline{\underline{X}}-\underline{\underline{\mathcal{I}}}\right)\frac{\rmd\hat{\Lambda}^{\left(u\right)}}{\rmd\underline{\underline{X}}}\left(\underline{\underline{X}}-\underline{\underline{\mathcal{I}}}\right).\label{eq:RelDerGOYX}
\end{equation}
On substituting the above identities in (\ref{eq:DifIndYX}) and on
simplifying terms we get:
\begin{eqnarray}
\fl\Biggl\{\hat{\underline{\gamma}}:\hat{\underline{\gamma}}^{T}\hat{\Lambda}:\Biggr\} & = & \rmi N\left(\underline{\underline{X}}\right)\left[2\underline{\underline{{\cal I}}}\hat{\Lambda}^{\left(u\right)}\left(\underline{\underline{X}}\right)+\left(\underline{\underline{X}}+\underline{\underline{{\cal I}}}\right)\frac{d\hat{\Lambda}^{\left(u\right)}\left(\underline{\underline{X}}\right)}{d\underline{\underline{X}}}\left(\underline{\underline{X}}-\underline{\underline{\mathcal{I}}}\right)\right].\label{eq:DifIdXUnGO}
\end{eqnarray}
We now wish to relate the derivative of the un-normalized Gaussian
operator with the derivative of the normalized one. Hence we use the
following relation:
\begin{eqnarray}
\fl\frac{\rmd\hat{\Lambda}\left(\underline{\underline{X}}\right)}{\rmd\underline{\underline{X}}} & = & \frac{\rmd}{\rmd\underline{\underline{X}}}N\left(\underline{\underline{X}}\right)\hat{\Lambda}^{\left(u\right)}\left(\underline{\underline{X}}\right)=N\left(\underline{\underline{X}}\right)\hat{\Lambda}^{\left(u\right)}\left(\underline{\underline{X}}\right)\left(\underline{\underline{X}}-\underline{\underline{\mathcal{I}}}\right)^{-1}+N\left(\underline{\underline{X}}\right)\frac{\rmd\hat{\Lambda}^{\left(u\right)}\left(\underline{\underline{X}}\right)}{\rmd\underline{\underline{X}}}.\label{eq:RelDer}
\end{eqnarray}
Here we have used the following result for the derivative of the square
root of the determinant of an antisymmetric matrix:
\[
\fl\frac{\rmd N\left(\underline{\underline{X}}\right)}{\rmd\underline{\underline{X}}}=\frac{\rmd}{\rmd\underline{\underline{X}}}\frac{1}{2^{M}}\sqrt{\det\left[\underline{\underline{X}}\right]}=\frac{1}{2^{M}}\sqrt{\det\left[\underline{\underline{X}}\right]}\left(\underline{\underline{X}}-\underline{\underline{\mathcal{I}}}\right)^{-1}=N\left(\underline{\underline{X}}\right)\left(\underline{\underline{X}}-\underline{\underline{\mathcal{I}}}\right)^{-1}.
\]
From (\ref{eq:RelDer}) we get: 
\begin{equation}
\frac{\rmd\hat{\Lambda}^{\left(u\right)}\left(\underline{\underline{X}}\right)}{\rmd\underline{\underline{X}}}=\frac{1}{N\left(\underline{\underline{X}}\right)}\frac{\rmd\hat{\Lambda}\left(\underline{\underline{X}}\right)}{\rmd\underline{\underline{X}}}-\hat{\Lambda}^{\left(u\right)}\left(\underline{\underline{X}}\right)\left(\underline{\underline{X}}-\underline{\underline{\mathcal{I}}}\right)^{-1}.\label{eq:RelDersNUn}
\end{equation}
On substituting this result on (\ref{eq:DifIdXUnGO}) and simplifying
we get:
\begin{eqnarray*}
\Biggl\{\hat{\underline{\gamma}}:\hat{\underline{\gamma}}^{T}\hat{\Lambda}:\Biggr\} & = & \rmi\left(\underline{\underline{X}}+\underline{\underline{{\cal I}}}\right)\frac{\rmd\hat{\Lambda}\left(\underline{\underline{X}}\right)}{\rmd\underline{\underline{X}}}\left(\underline{\underline{X}}-\underline{\underline{\mathcal{I}}}\right)-\rmi\left(\underline{\underline{X}}-\underline{\underline{{\cal I}}}\right)\hat{\Lambda}\left(\underline{\underline{X}}\right).
\end{eqnarray*}
Therefore we have proved the differential identity given in (\ref{eq:MajMixProd}).

\subsection{Second mixed product}

\[
:\hat{\underline{\gamma}}\Biggl\{\hat{\underline{\gamma}}^{T}\hat{\Lambda}\Biggr\}:=\rmi\left[\underline{\underline{X}}^{-}\frac{\rmd\hat{\Lambda}}{\rmd\underline{\underline{X}}}-\hat{\Lambda}\right]\underline{\underline{X}}^{+}
\]

\textbf{\emph{Proof.}} We can derive this fourth mixed identity from
the third one. We know the orderings of both identities as given in
(\ref{eq:ExpFMP} and \ref{eq:ExpSMP}). Utilizing those orderings
and Fermi commutation relation in (\ref{eq:FerComm}) we get:

\[
:\hat{\underline{\gamma}}\Biggl\{\hat{\underline{\gamma}}^{T}\hat{\Lambda}\Biggr\}:=-\Bigl\{\hat{\underline{\gamma}}:\hat{\underline{\gamma}}^{T}\hat{\Lambda}:\Bigr\}^{T}-2i\hat{\Lambda}\underline{\underline{{\cal I}}}
\]

This will give us the new fourth identity as in (\ref{eq:Majmix2})

\subsection{Normally ordered products }

\[
:\hat{\underline{\gamma}}\hat{\underline{\gamma}}^{T}\hat{\Lambda}:=\rmi\left(\underline{\underline{X}}-\underline{\underline{\mathcal{I}}}\right)\hat{\Lambda}\left(\underline{\underline{X}}\right)-\rmi\left(\underline{\underline{X}}-\underline{\underline{\mathcal{I}}}\right)\frac{\rmd\hat{\Lambda}\left(\underline{\underline{X}}\right)}{\rmd\underline{\underline{X}}}\left(\underline{\underline{X}}-\underline{\underline{\mathcal{I}}}\right).
\]
\textbf{Proof}. We use $\hat{\Lambda}\left(\underline{\underline{X}}\right)=N\left(\underline{\underline{X}}\right)\hat{\Lambda}^{\left(u\right)}\left(\underline{\underline{X}}\right)$
and we also use the expressions given in (\ref{eq:RelDerGOYX}) and
(\ref{eq:RelDersNUn}) in order to express the derivative in terms
of the normalized Gaussian operators which is 
\begin{eqnarray}
\fl\frac{\rmd}{\rmd\underline{\underline{Y}}}\hat{\Lambda}^{\left(u\right)} & = & -\frac{1}{N\left(\underline{\underline{X}}\right)}\left(\underline{\underline{X}}-\underline{\underline{\mathcal{I}}}\right)\frac{\rmd\hat{\Lambda}\left(\underline{\underline{X}}\right)}{\rmd\underline{\underline{X}}}\left(\underline{\underline{X}}-\underline{\underline{\mathcal{I}}}\right)+\left(\underline{\underline{X}}-\underline{\underline{\mathcal{I}}}\right)\hat{\Lambda}^{\left(u\right)}\left(\underline{\underline{X}}\right).\label{eq:derUnNGOXY}
\end{eqnarray}
On substituting the above result in (\ref{eq:UnNormNormalDifId})
we get:
\begin{eqnarray*}
:\hat{\underline{\gamma}}\hat{\underline{\gamma}}^{T}\hat{\Lambda}: & = & \rmi\left(\underline{\underline{X}}-\underline{\underline{\mathcal{I}}}\right)\hat{\Lambda}\left(\underline{\underline{X}}\right)-\rmi\left(\underline{\underline{X}}-\underline{\underline{\mathcal{I}}}\right)\frac{\rmd\hat{\Lambda}\left(\underline{\underline{X}}\right)}{\rmd\underline{\underline{X}}}\left(\underline{\underline{X}}-\underline{\underline{\mathcal{I}}}\right).
\end{eqnarray*}
Thus we have proved the differential identity given in (\ref{eq:NormalNormMajId}).

\subsection{Anti-normally ordered products }

\[
\left\{ \underline{\widehat{\gamma}}\widehat{\underline{\gamma}}^{T}\hat{\Lambda}\right\} =\rmi\left[-\left(\underline{\underline{\mathcal{I}}}+\underline{\underline{X}}\right)\frac{\rmd\hat{\Lambda}\left(\underline{\underline{X}}\right)}{\rmd\underline{\underline{X}}}\left(\underline{\underline{\mathcal{I}}}+\underline{\underline{X}}\right)+\hat{\Lambda}\left(\underline{\underline{X}}\right)\left(\underline{\underline{\mathcal{I}}}+\underline{\underline{X}}\right)\right].
\]
\textbf{Proof}. As in the previous cases we use $\hat{\Lambda}\left(\underline{\underline{X}}\right)=N\left(\underline{\underline{X}}\right)\hat{\Lambda}^{\left(u\right)}\left(\underline{\underline{X}}\right)$
and we also make a change of variables using (\ref{eq:ChangeVarYX})
on the differential identities given in (\ref{eq:AntiNormDifIdUnGO})
obtaining: 
\[
\fl\left\{ \underline{\widehat{\gamma}}\widehat{\underline{\gamma}}^{T}\hat{\Lambda}\right\} =\rmi N\left(\underline{\underline{X}}\right)\underline{\underline{\mathcal{I}}}\left\{ \left(2\left(\underline{\underline{X}}-\underline{\underline{\mathcal{I}}}\right)^{-1}-\underline{\underline{\mathcal{I}}}\right)\frac{\rmd}{\rmd\underline{\underline{Y}}}-2\right\} \hat{\Lambda}^{\left(u\right)}\left(\underline{\underline{X}}\right)\left(2\left(\underline{\underline{X}}-\underline{\underline{\mathcal{I}}}\right)^{-1}-\underline{\underline{\mathcal{I}}}\right)\underline{\underline{\mathcal{I}}}.
\]
 Next, we express the derivative in terms of the Gaussian operators
and the $\underline{\underline{X}}$ matrix using the identity given
in (\ref{eq:derUnNGOXY}) obtaining:
\begin{eqnarray*}
 &  & \fl\left\{ \underline{\widehat{\gamma}}\widehat{\underline{\gamma}}^{T}\hat{\Lambda}\right\} =\rmi\underline{\underline{\mathcal{I}}}\left\{ \left(2\left(\underline{\underline{X}}-\underline{\underline{\mathcal{I}}}\right)^{-1}-\underline{\underline{\mathcal{I}}}\right)\left[-\left(\underline{\underline{X}}-\underline{\underline{\mathcal{I}}}\right)\frac{\rmd\hat{\Lambda}\left(\underline{\underline{X}}\right)}{\rmd\underline{\underline{X}}}\left(\underline{\underline{X}}-\underline{\underline{\mathcal{I}}}\right)+\left(\underline{\underline{X}}-\underline{\underline{\mathcal{I}}}\right)\hat{\Lambda}\left(\underline{\underline{X}}\right)\right]\right.\\
 &  & \left.-2\hat{\Lambda}\left(\underline{\underline{X}}\right)\right\} \left(2\left(\underline{\underline{X}}-\underline{\underline{\mathcal{I}}}\right)^{-1}-\underline{\underline{\mathcal{I}}}\right)\underline{\underline{\mathcal{I}}}.
\end{eqnarray*}
On simplifying terms and using that $\underline{\underline{\mathcal{I}}}\mathbf{\underline{\underline{\mathcal{I}}}}=-\underline{\underline{I}}$
we get:
\begin{eqnarray*}
\left\{ \underline{\widehat{\gamma}}\widehat{\underline{\gamma}}^{T}\hat{\Lambda}\right\}  & = & \rmi\left[-\left(\underline{\underline{\mathcal{I}}}+\underline{\underline{X}}\right)\frac{\rmd\hat{\Lambda}\left(\underline{\underline{X}}\right)}{\rmd\underline{\underline{X}}}\left(\underline{\underline{\mathcal{I}}}+\underline{\underline{X}}\right)+\hat{\Lambda}\left(\underline{\underline{X}}\right)\left(\underline{\underline{\mathcal{I}}}+\underline{\underline{X}}\right)\right].
\end{eqnarray*}
Therefore we have proved the differential identity given in (\ref{eq:AntiNormalDifId}).

\section{Majorana differential identities for unordered products\label{sec:AppendixDiffIdunorderedGO-1}}

In this Section we give the detailed calculations for unordered Majorana
differential identities. These are derived using the ordered identities
given in \ref{sec:AppendixDiffIdNormGO}.

\subsection{Unordered left and right products}

\begin{equation}
\hat{\underline{\gamma}}\hat{\underline{\gamma}}^{T}\hat{\Lambda}=\rmi\left[\left(1+\rmi\underline{\underline{x}}\right)\frac{d\hat{\Lambda}}{d\underline{\underline{x}}}\left(1-\rmi\underline{\underline{x}}\right)-\hat{\Lambda}\left(\underline{\underline{x}}+\rmi\right)\right].\label{eq:Un-OrdLP}
\end{equation}

\textbf{Proof}. The explicit expression for $\hat{\underline{\gamma}}\hat{\underline{\gamma}}^{T}\hat{\Lambda}$
in terms of the Fermi operators is given in (\ref{eq:ExpUn-Or}).
The idea is to find the appropriate combination between the four different
expressions of the ordered products of Majorana operators and the
Gaussian operator that gives the expression given in (\ref{eq:ExpUn-Or}).
Each ordered product will be used in order to obtain the products
of only Fermi creation operators with the Gaussian operators, or Fermi
annihilation operators and the Gaussian operators or the two different
product of Fermi creation and annihilation operator with the Gaussian
operator in the left side. In order to do this we will use the real
antisymmetric matrix $\underline{\underline{{\cal I}}}$. We first
obtain the combination of the anti-normal product that will give the
matrix form with only product of Fermi annihilation operators and
the Gaussian one. Hence on pre and post multiplying (\ref{eq:ExpANOP})
by $\underline{\underline{{\cal I}}}$, we get:
\begin{eqnarray}
 &  & \fl\underline{\underline{{\cal I}}}\left\{ \hat{\gamma}_{\mu}\hat{\gamma}_{\upsilon}\hat{\Lambda}\right\} \underline{\underline{{\cal I}}}=\nonumber \\
 &  & \fl\left(\begin{array}{cc}
\widehat{a}_{i}\widehat{a}_{j}\hat{\Lambda}+\hat{\Lambda}\widehat{a}_{i}^{\dagger}\widehat{a}_{j}^{\dagger}-\widehat{a}_{i}\hat{\Lambda}\widehat{a}_{j}^{\dagger}+\widehat{a}_{j}\hat{\Lambda}\widehat{a}_{i}^{\dagger} & -i\widehat{a}_{i}\widehat{a}_{j}\hat{\Lambda}+i\hat{\Lambda}\widehat{a}_{i}^{\dagger}\widehat{a}_{j}^{\dagger}-i\widehat{a}_{i}\hat{\Lambda}\widehat{a}_{j}^{\dagger}-i\widehat{a}_{j}\hat{\Lambda}\widehat{a}_{i}^{\dagger}\\
-i\widehat{a}_{i}\widehat{a}_{j}\hat{\Lambda}+i\hat{\Lambda}\widehat{a}_{i}^{\dagger}\widehat{a}_{j}^{\dagger}+i\widehat{a}_{i}\hat{\Lambda}\widehat{a}_{j}^{\dagger}+i\widehat{a}_{j}\hat{\Lambda}\widehat{a}_{i}^{\dagger} & -\widehat{a}_{i}\widehat{a}_{j}\hat{\Lambda}-\hat{\Lambda}\widehat{a}_{i}^{\dagger}\widehat{a}_{j}^{\dagger}-\widehat{a}_{i}\hat{\Lambda}\widehat{a}_{j}^{\dagger}+\widehat{a}_{j}\hat{\Lambda}\widehat{a}_{i}^{\dagger}
\end{array}\right).\label{eq:IpreIpost}
\end{eqnarray}
Now adding (\ref{eq:ExpANOP}) and (\ref{eq:IpreIpost}) leads to:

\begin{eqnarray}
 &  & B_{1}=\left(\begin{array}{cc}
2\widehat{a}_{i}\widehat{a}_{j}\hat{\Lambda}+2\hat{\Lambda}\widehat{a}_{i}^{\dagger}\widehat{a}_{j}^{\dagger} & -2\rmi\widehat{a}_{i}\widehat{a}_{j}\hat{\Lambda}+2\rmi\hat{\Lambda}\widehat{a}_{i}^{\dagger}\widehat{a}_{j}^{\dagger}\\
-2\rmi\widehat{a}_{i}\widehat{a}_{j}\hat{\Lambda}+2\rmi\hat{\Lambda}\widehat{a}_{i}^{\dagger}\widehat{a}_{j}^{\dagger} & -2\widehat{a}_{i}\widehat{a}_{j}\hat{\Lambda}-2\hat{\Lambda}\widehat{a}_{i}^{\dagger}\widehat{a}_{j}^{\dagger}
\end{array}\right),\label{eq:B1}
\end{eqnarray}
where $B_{1}=\left\{ \hat{\gamma}_{\mu}\hat{\gamma}_{\upsilon}\hat{\Lambda}\right\} +\underline{\underline{{\cal I}}}\left\{ \hat{\gamma}_{\mu}\hat{\gamma}_{\upsilon}\hat{\Lambda}\right\} \underline{\underline{{\cal I}}}.$
We notice that we can obtain the required product of Fermi annihilation
operators and $\hat{\Lambda}$ from the following expression:

\begin{eqnarray}
\frac{B_{1}-\rmi B_{1}\underline{\underline{{\cal I}}}}{4} & = & \left(\begin{array}{cc}
\widehat{a}_{i}\widehat{a}_{j}\hat{\Lambda} & -i\widehat{a}_{i}\widehat{a}_{j}\hat{\Lambda}\\
-i\widehat{a}_{i}\widehat{a}_{j}\hat{\Lambda} & -\widehat{a}_{i}\widehat{a}_{j}\hat{\Lambda}
\end{array}\right).\label{eq:antinormal_part}
\end{eqnarray}
In a similar way we wish to obtain a matrix of the product of Fermi
creation operators with $\hat{\Lambda}$. In this case we consider
(\ref{eq:ENOP}) and perform the same procedures as described above
obtaining:
\[
B_{2}=\left(\begin{array}{cc}
2\widehat{a}_{i}^{\dagger}\widehat{a}_{j}^{\dagger}\hat{\Lambda}+2\hat{\Lambda}\widehat{a}_{i}\widehat{a}_{j} & -\rmi\left[-2\widehat{a}_{i}^{\dagger}\widehat{a}_{j}^{\dagger}\hat{\Lambda}+2\hat{\Lambda}\widehat{a}_{i}\widehat{a}_{j}\right]\\
-\rmi\left[-2\widehat{a}_{i}^{\dagger}\widehat{a}_{j}^{\dagger}\hat{\Lambda}+2\hat{\Lambda}\widehat{a}_{i}\widehat{a}_{j}\right] & -2\widehat{a}_{i}^{\dagger}\widehat{a}_{j}^{\dagger}\hat{\Lambda}-2\hat{\Lambda}\widehat{a}_{i}\widehat{a}_{j}
\end{array}\right),
\]
where $B_{2}=:\hat{\gamma}_{\mu}\hat{\gamma}_{\upsilon}\hat{\Lambda}:+\underline{\underline{{\cal I}}}:\hat{\gamma}_{\mu}\hat{\gamma}_{\upsilon}\hat{\Lambda}:\underline{\underline{{\cal I}}}$.
Then the required expression is:

\begin{eqnarray}
\frac{B_{2}+\rmi B_{2}\underline{\underline{{\cal I}}}}{4} & = & \left(\begin{array}{cc}
\widehat{a}_{i}^{\dagger}\widehat{a}_{j}^{\dagger}\hat{\Lambda} & \rmi\widehat{a}_{i}^{\dagger}\widehat{a}_{j}^{\dagger}\hat{\Lambda}\\
\rmi\widehat{a}_{i}^{\dagger}\widehat{a}_{j}^{\dagger}\hat{\Lambda} & -\widehat{a}_{i}^{\dagger}\widehat{a}_{j}^{\dagger}\hat{\Lambda}
\end{array}\right).\label{eq:Normal_part}
\end{eqnarray}
Similarly, we use the mixed product given in (\ref{eq:ExpFMP}) in
order to obtain:
\[
C_{1}=\left(\begin{array}{cc}
2\widehat{a}_{i}\widehat{a}_{j}^{\dagger}\hat{\Lambda}-2\hat{\Lambda}\widehat{a}_{j}\widehat{a}_{i}^{\dagger} & \rmi\left[2\widehat{a}_{i}\widehat{a}_{j}^{\dagger}\hat{\Lambda}+2\hat{\Lambda}\widehat{a}_{j}\widehat{a}_{i}^{\dagger}\right]\\
\rmi\left[-2\widehat{a}_{i}\widehat{a}_{j}^{\dagger}\hat{\Lambda}-2\hat{\Lambda}\widehat{a}_{j}\widehat{a}_{i}^{\dagger}\right] & 2\widehat{a}_{i}\widehat{a}_{j}^{\dagger}\widehat{\Lambda}-2\widehat{\Lambda}\widehat{a}_{j}\widehat{a}_{i}^{\dagger}
\end{array}\right),
\]
where $C_{1}=\Bigl\{\hat{\gamma}_{\mu}:\hat{\gamma}_{\upsilon}\hat{\Lambda}:\Bigr\}-\underline{\underline{{\cal I}}}\Bigl\{\hat{\gamma}_{\mu}:\hat{\gamma}_{\upsilon}\hat{\Lambda}:\Bigr\}\underline{\underline{{\cal I}}}$
and from this expression we get a matrix in terms of product of creation
and annihilation operators and $\hat{\Lambda}$:

\begin{eqnarray}
\frac{C_{1}+\rmi C_{1}\underline{\underline{{\cal I}}}}{4} & = & \left(\begin{array}{cc}
\widehat{a}_{i}\widehat{a}_{j}^{\dagger}\hat{\Lambda} & \rmi\widehat{a}_{i}\widehat{a}_{j}^{\dagger}\hat{\Lambda}\\
-\rmi\widehat{a}_{i}\widehat{a}_{j}^{\dagger}\hat{\Lambda} & \widehat{a}_{i}\widehat{a}_{j}^{\dagger}\widehat{\Lambda}
\end{array}\right).\label{eq:mixedpart1}
\end{eqnarray}
Following the same steps as described above from (\ref{eq:ExpSMP})
we get: 
\[
C_{2}=\left(\begin{array}{cc}
2\widehat{a}_{i}^{\dagger}\widehat{a}_{j}\hat{\Lambda}-2\hat{\Lambda}\widehat{a}_{j}^{\dagger}\widehat{a}_{i} & \rmi\left[-2\widehat{a}_{i}^{\dagger}\widehat{a}_{j}\hat{\Lambda}-2\hat{\Lambda}\widehat{a}_{j}^{\dagger}\widehat{a}_{i}\right]\\
\rmi\left[+2\widehat{a}_{i}^{\dagger}\widehat{a}_{j}\hat{\Lambda}+2\hat{\Lambda}\widehat{a}_{j}^{\dagger}\widehat{a}_{i}\right] & 2\widehat{a}_{i}^{\dagger}\widehat{a}_{j}\hat{\Lambda}-2\hat{\Lambda}\widehat{a}_{j}^{\dagger}\widehat{a}_{i}
\end{array}\right),
\]
where: $C_{2}=:\hat{\gamma}_{\mu}\Biggl\{\hat{\gamma}_{\upsilon}\hat{\Lambda}\Biggr\}:-\underline{\underline{{\cal I}}}:\hat{\gamma}_{\mu}\Biggl\{\hat{\gamma}_{\upsilon}\hat{\Lambda}\Biggr\}:\underline{\underline{{\cal I}}}$
and we also obtain:
\begin{eqnarray}
\frac{C_{2}-\rmi C_{2}\underline{\underline{{\cal I}}}}{4} & = & \left(\begin{array}{cc}
\widehat{a}_{i}^{\dagger}\widehat{a}_{j}\hat{\Lambda} & -i\widehat{a}_{i}^{\dagger}\widehat{a}_{j}\hat{\Lambda}\\
i\widehat{a}_{i}^{\dagger}\widehat{a}_{j}\hat{\Lambda} & \widehat{a}_{i}^{\dagger}\widehat{a}_{j}\hat{\Lambda}
\end{array}\right).\label{eq:mixed part2}
\end{eqnarray}
On adding the identities given in (\ref{eq:Normal_part}), (\ref{eq:antinormal_part}),
(\ref{eq:mixedpart1}), and (\ref{eq:mixed part2}) we obtain the
combination of Fermi operators and the Gaussian operators corresponding
to the unordered product of Majorana operators and $\hat{\Lambda}$
given in (\ref{eq:ExpUn-Or}): 

\[
\hat{\underline{\gamma}}\hat{\underline{\gamma}}^{T}\hat{\Lambda}=\frac{C_{1}+\rmi C_{1}\underline{\underline{{\cal I}}}}{4}+\frac{C_{2}-\rmi C_{2}\underline{\underline{{\cal I}}}}{4}+\frac{B_{1}-\rmi B_{1}\underline{\underline{{\cal I}}}}{4}+\frac{B_{2}+\rmi B_{2}\underline{\underline{{\cal I}}}}{4}.
\]
In order to obtain the differential identity we notice that $B_{1},$
$B_{2}$, $C_{1}$ and $C_{2}$ and given in terms of the normalized
ordered differential identities derived in \ref{sec:AppendixDiffIdNormGO}.
Thus using the expression in the right hand side of the four normalized
differential identities and on simplifying terms we get:

\[
\hat{\underline{\gamma}}\hat{\underline{\gamma}}^{T}\hat{\Lambda}=-\rmi\underline{\underline{{\cal I}}}\frac{d\hat{\Lambda}}{d\underline{\underline{X}}}\underline{\underline{{\cal I}}}-\rmi\underline{\underline{{\cal I}}}\underline{\underline{X}}\frac{d\hat{\Lambda}}{d\underline{\underline{X}}}\underline{\underline{X}}\underline{\underline{{\cal I}}}+\rmi\underline{\underline{{\cal I}}}\hat{\Lambda}\underline{\underline{X}}\underline{\underline{{\cal I}}}-\underline{\underline{{\cal I}}}\frac{d\hat{\Lambda}}{d\underline{\underline{X}}}\underline{\underline{X}}\underline{\underline{{\cal I}}}+\underline{\underline{{\cal I}}}\underline{\underline{X}}\frac{d\hat{\Lambda}}{d\underline{\underline{X}}}\underline{\underline{{\cal I}}}+\hat{\Lambda}\underline{\underline{I}}.
\]
The above equation can be written in a simple form as:

\begin{equation}
\hat{\underline{\gamma}}\hat{\underline{\gamma}}^{T}\hat{\Lambda}=-\rmi\underline{\underline{{\cal I}}}\left[\left(1+i\underline{\underline{X}}\right)\frac{d\hat{\Lambda}}{d\underline{\underline{X}}}\left(1-\rmi\underline{\underline{X}}\right)-\rmi\hat{\Lambda}\left(1-\rmi\underline{\underline{X}}\right)\right]\underline{\underline{{\cal I}}}.\label{eq:id}
\end{equation}
In terms of the alternate antisymmetric form, $\underline{\underline{x}}=\underline{\underline{{\cal I}}}\underline{\underline{X}}^{T}\underline{\underline{{\cal I}}}$
, we can define:
\begin{equation}
\frac{d\hat{\Lambda}}{d\underline{\underline{x}}^{T}}=\underline{\underline{{\cal I}}}\frac{d\hat{\Lambda}}{d\underline{\underline{X}}}\underline{\underline{{\cal I}}}\,.\label{eq:dLdxT}
\end{equation}
In this way we can introduce $\underline{\underline{x}}^{\pm}=\underline{\underline{x}}\pm\rmi\underline{\underline{I}},$
and simplify (\ref{eq:id}) to give:
\begin{equation}
\hat{\underline{\gamma}}\hat{\underline{\gamma}}^{T}\hat{\Lambda}=\rmi\left[\underline{\underline{x}}^{-}\frac{\rmd\hat{\Lambda}}{\rmd\underline{\underline{x}}}\underline{\underline{x}}^{+}-\hat{\Lambda}\underline{\underline{x}}^{+}\right].
\end{equation}
On conjugating this expression, we obtain the right product identity:
\begin{equation}
\hat{\Lambda}\hat{\underline{\gamma}}\hat{\underline{\gamma}}^{T}=\rmi\left[\underline{\underline{x}}^{+}\frac{\rmd\hat{\Lambda}}{\rmd\underline{\underline{x}}}\underline{\underline{x}}^{-}-\hat{\Lambda}\underline{\underline{x}}^{-}\right].\label{eq:UnorderedDifId-1-1}
\end{equation}

\subsection{Unordered mixed products}

\[
\hat{\underline{\gamma}}\hat{\Lambda}\hat{\underline{\gamma}}^{T}=\rmi\left[\left(i\underline{\underline{x}}+\underline{\underline{I}}\right)\frac{d\hat{\Lambda}}{d\underline{\underline{x}}}\left(\underline{\underline{I}}+\rmi\underline{\underline{x}}\right)-i\hat{\Lambda}\left(\underline{\underline{I}}+\rmi\underline{\underline{x}}\right)\right].
\]

\textbf{Proof}. Analogous to the previous differential identity, the
method is to use the four different orderings of products of Majorana
operators and Gaussian operators to give an explicit expression for
$\hat{\underline{\gamma}}\hat{\Lambda}\hat{\underline{\gamma}}^{T}$,
in terms of the ordered Fermi operators and Gaussian operators. Next
we transform this into the corresponding differential identities. 

First, we obtain matrices of products of the Fermi creation and annihilation
operators where the Gaussian operator is in the middle. Thus, pre
and post multiplying (\ref{eq:ExpANOP}) by the real antisymmetric
matrix $\underline{\underline{{\cal I}}}$, we get:

\begin{eqnarray}
 &  & \fl\underline{\underline{{\cal I}}}\left\{ \hat{\gamma}_{\mu}\hat{\gamma}_{\upsilon}\hat{\Lambda}\right\} \underline{\underline{{\cal I}}}=\nonumber \\
 &  & \fl\left(\begin{array}{cc}
\widehat{a}_{i}\widehat{a}_{j}\hat{\Lambda}+\hat{\Lambda}\widehat{a}_{i}^{\dagger}\widehat{a}_{j}^{\dagger}-\widehat{a}_{i}\hat{\Lambda}\widehat{a}_{j}^{\dagger}+\widehat{a}_{j}\hat{\Lambda}\widehat{a}_{i}^{\dagger} & -i\widehat{a}_{i}\widehat{a}_{j}\hat{\Lambda}+i\hat{\Lambda}\widehat{a}_{i}^{\dagger}\widehat{a}_{j}^{\dagger}-i\widehat{a}_{i}\hat{\Lambda}\widehat{a}_{j}^{\dagger}-i\widehat{a}_{j}\hat{\Lambda}\widehat{a}_{i}^{\dagger}\\
-i\widehat{a}_{i}\widehat{a}_{j}\hat{\Lambda}+i\hat{\Lambda}\widehat{a}_{i}^{\dagger}\widehat{a}_{j}^{\dagger}+i\widehat{a}_{i}\hat{\Lambda}\widehat{a}_{j}^{\dagger}+i\widehat{a}_{j}\hat{\Lambda}\widehat{a}_{i}^{\dagger} & -\widehat{a}_{i}\widehat{a}_{j}\hat{\Lambda}-\hat{\Lambda}\widehat{a}_{i}^{\dagger}\widehat{a}_{j}^{\dagger}-\widehat{a}_{i}\hat{\Lambda}\widehat{a}_{j}^{\dagger}+\widehat{a}_{j}\hat{\Lambda}\widehat{a}_{i}^{\dagger}
\end{array}\right),\label{eq:IpreIpost-1}
\end{eqnarray}
Adding (\ref{eq:ExpANOP}) and (\ref{eq:IpreIpost-1}) leads to:

\begin{eqnarray}
D_{1} & = & \left(\begin{array}{cc}
2\widehat{a}_{i}\hat{\Lambda}\widehat{a}_{j}^{\dagger}-2\widehat{a}_{j}\hat{\Lambda}\widehat{a}_{i}^{\dagger} & 2i\widehat{a}_{i}\hat{\Lambda}\widehat{a}_{j}^{\dagger}+2i\widehat{a}_{j}\hat{\Lambda}\widehat{a}_{i}^{\dagger}\\
-2i\widehat{a}_{i}\hat{\Lambda}\widehat{a}_{j}^{\dagger}-2i\widehat{a}_{j}\hat{\Lambda}\widehat{a}_{i}^{\dagger} & 2\widehat{a}_{i}\hat{\Lambda}\widehat{a}_{j}^{\dagger}-2\widehat{a}_{j}\hat{\Lambda}\widehat{a}_{i}^{\dagger}
\end{array}\right),\label{eq:B1-1}
\end{eqnarray}
where $D_{1}=\left\{ \hat{\gamma}_{\mu}\hat{\gamma}_{\upsilon}\hat{\Lambda}\right\} -\underline{\underline{{\cal I}}}\left\{ \hat{\gamma}_{\mu}\hat{\gamma}_{\upsilon}\hat{\Lambda}\right\} \underline{\underline{{\cal I}}}$.
Using this expression we obtain:
\begin{eqnarray}
\frac{D_{1}+\rmi D_{1}\underline{\underline{{\cal I}}}}{4} & = & \left(\begin{array}{cc}
\widehat{a}_{i}\hat{\Lambda}\widehat{a}_{j}^{\dagger} & \rmi\widehat{a}_{i}\hat{\Lambda}\widehat{a}_{j}^{\dagger}\\
-\rmi\widehat{a}_{i}\hat{\Lambda}\widehat{a}_{j}^{\dagger} & \widehat{a}_{i}\hat{\Lambda}\widehat{a}_{j}^{\dagger}
\end{array}\right).\label{eq:antinormal_part-1}
\end{eqnarray}

In this case, using the anti-normal ordering identity we obtain a
matrix with products of the form $\widehat{a}_{i}\hat{\Lambda}\widehat{a}_{j}^{\dagger}$.
We now use the the same procedure as described above for the normal
ordering product, (\ref{eq:ENOP}), obtaining:
\[
D_{2}=\left(\begin{array}{cc}
-2\widehat{a}_{j}^{\dagger}\hat{\Lambda}\widehat{a}_{i}+2\widehat{a}_{i}^{\dagger}\hat{\Lambda}\widehat{a}_{j} & -i2\widehat{a}_{j}^{\dagger}\hat{\Lambda}\widehat{a}_{i}-i2\widehat{a}_{i}^{\dagger}\hat{\Lambda}\widehat{a}_{j}\\
i2\widehat{a}_{j}^{\dagger}\hat{\Lambda}\widehat{a}_{i}+i2\widehat{a}_{i}^{\dagger}\hat{\Lambda}\widehat{a}_{j} & -2\widehat{a}_{j}^{\dagger}\hat{\Lambda}\widehat{a}_{i}+2\widehat{a}_{i}^{\dagger}\hat{\Lambda}\widehat{a}_{j}
\end{array}\right),
\]
where $D_{2}=:\hat{\gamma}_{\mu}\hat{\gamma}_{\upsilon}\hat{\Lambda}:-\underline{\underline{{\cal I}}}:\hat{\gamma}_{\mu}\hat{\gamma}_{\upsilon}\hat{\Lambda}:\underline{\underline{{\cal I}}}$
. From this we get a matrix with products of the form $\widehat{a}_{i}^{\dagger}\hat{\Lambda}\widehat{a}_{j}$:

\begin{eqnarray}
\frac{D_{2}-\rmi D_{2}\underline{\underline{{\cal I}}}}{4} & = & \left(\begin{array}{cc}
\widehat{a}_{i}^{\dagger}\hat{\Lambda}\widehat{a}_{j} & -\rmi\widehat{a}_{i}^{\dagger}\hat{\Lambda}\widehat{a}_{j}\\
\rmi\widehat{a}_{i}^{\dagger}\hat{\Lambda}\widehat{a}_{j} & \widehat{a}_{i}^{\dagger}\hat{\Lambda}\widehat{a}_{j}
\end{array}\right).\label{eq:Normal_part-1}
\end{eqnarray}
Next, using (\ref{eq:ExpFMP}) we arrive at the following expression:

\begin{eqnarray}
\frac{E_{1}-\rmi E_{1}\underline{\underline{{\cal I}}}}{4} & = & \left(\begin{array}{cc}
\widehat{a}_{i}\hat{\Lambda}\widehat{a}_{j} & -\rmi\widehat{a}_{i}\hat{\Lambda}\widehat{a}_{j}\\
-\rmi\widehat{a}_{i}\hat{\Lambda}\widehat{a}_{j} & -\widehat{a}_{i}\widehat{\Lambda}\widehat{a}_{j}
\end{array}\right),\label{eq:mixedpart1-1}
\end{eqnarray}
where $E_{1}=\Bigl\{\hat{\gamma}_{\mu}:\hat{\gamma}_{\upsilon}\hat{\Lambda}:\Bigr\}+\underline{\underline{{\cal I}}}\Bigl\{\hat{\gamma}_{\mu}:\hat{\gamma}_{\upsilon}\hat{\Lambda}:\Bigr\}\underline{\underline{{\cal I}}}$. 

Using (\ref{eq:ExpSMP}) we obtain a matrix with products of the form
$\widehat{a}_{i}^{\dagger}\hat{\Lambda}\widehat{a}_{j}^{\dagger}$:

\begin{eqnarray}
\frac{E_{2}+\rmi E_{2}\underline{\underline{{\cal I}}}}{4} & = & \left(\begin{array}{cc}
\widehat{a}_{i}^{\dagger}\hat{\Lambda}\widehat{a}_{j}^{\dagger} & \rmi\widehat{a}_{i}^{\dagger}\hat{\Lambda}\widehat{a}_{j}^{\dagger}\\
\rmi\widehat{a}_{i}^{\dagger}\hat{\Lambda}\widehat{a}_{j}^{\dagger} & -\widehat{a}_{i}^{\dagger}\hat{\Lambda}\widehat{a}_{j}^{\dagger}
\end{array}\right),\label{eq:mixed part2-1}
\end{eqnarray}
where: $E_{2}=:\hat{\gamma}_{\mu}\Biggl\{\hat{\gamma}_{\upsilon}\hat{\Lambda}\Biggr\}:+\underline{\underline{{\cal I}}}:\hat{\gamma}_{\mu}\Biggl\{\hat{\gamma}_{\upsilon}\hat{\Lambda}\Biggr\}:\underline{\underline{{\cal I}}}$.

On adding the expressions given in (\ref{eq:Normal_part-1}), (\ref{eq:antinormal_part-1}),
(\ref{eq:mixedpart1-1}), and (\ref{eq:mixed part2-1}) we get the
appropriate combination of products of Fermi operators and the Gaussian
operator:

\[
\hat{\underline{\gamma}}\hat{\Lambda}\hat{\underline{\gamma}}^{T}=\frac{E_{1}-\rmi E_{1}\underline{\underline{{\cal I}}}}{4}+\frac{E_{2}+\rmi E_{2}\underline{\underline{{\cal I}}}}{4}+\frac{D_{1}+\rmi D_{1}\underline{\underline{{\cal I}}}}{4}+\frac{D_{2}-\rmi D_{2}\underline{\underline{{\cal I}}}}{4}.
\]
Now utilizing the right hand side of the four normalized differential
identities an on simplifying terms we arrive at the following expression:
\[
\hat{\underline{\gamma}}\hat{\Lambda}\hat{\underline{\gamma}}^{T}=-i\underline{\underline{{\cal I}}}\frac{d\hat{\Lambda}}{d\underline{\underline{X}}}\underline{\underline{{\cal I}}}+i\underline{\underline{{\cal I}}}\underline{\underline{X}}\frac{d\hat{\Lambda}}{d\underline{\underline{X}}}\underline{\underline{X}}\underline{\underline{{\cal I}}}-i\underline{\underline{{\cal I}}}\hat{\Lambda}\underline{\underline{X}}\underline{\underline{{\cal I}}}+\underline{\underline{{\cal I}}}\frac{d\hat{\Lambda}}{d\underline{\underline{X}}}\underline{\underline{X}}\underline{\underline{{\cal I}}}+\underline{\underline{{\cal I}}}\underline{\underline{X}}\frac{d\hat{\Lambda}}{d\underline{\underline{X}}}\underline{\underline{{\cal I}}}+\hat{\Lambda}\underline{\underline{I}}.
\]
Taking common terms outside we get:

\begin{equation}
\hat{\underline{\gamma}}\hat{\Lambda}\hat{\underline{\gamma}}^{T}=-i\underline{\underline{{\cal I}}}\left[\left(i\underline{\underline{X}}+1\right)\frac{d\hat{\Lambda}}{d\underline{\underline{X}}}\left(1+i\underline{\underline{X}}\right)-i\hat{\Lambda}\left(1+i\underline{\underline{X}}\right)\right]\underline{\underline{{\cal I}}}.\label{eq:id-1}
\end{equation}
This leads to our final result, which can also be written as:
\begin{equation}
\hat{\underline{\gamma}}\hat{\Lambda}\hat{\underline{\gamma}}^{T}=\rmi\left[-\underline{\underline{x}}^{-}\frac{\rmd\hat{\Lambda}}{\rmd\underline{\underline{x}}}\underline{\underline{x}}^{-}+\hat{\Lambda}\underline{\underline{x}}^{-}\right].
\end{equation}

\section{Time evolution of the Majorana Q-function\label{sec:AppendixTimeevolution}}

Here we provide the details of the calculations in order to derive
(\ref{eq:TEf}), (\ref{eq:TEf2}) and (\ref{eq:TEX}) of Section \ref{sec:Time-Evolution-MQf}.
The chain rule allow us to obtain the relation between the Gaussian
basis $\hat{\Lambda}^{N}\left(\underline{\underline{x}}\right)$ and
the Gaussian operator $\hat{\Lambda}\left(\underline{\underline{x}}\right),$
which is:
\begin{equation}
\frac{1}{{\cal N}}S\left(\left[\underline{\underline{x}}\right]^{2}\right)\frac{\rmd\hat{\Lambda}\left(\underline{\underline{x}}\right)}{\rmd\underline{\underline{x}}}=\frac{\rmd\hat{\Lambda}^{N}\left(\underline{\underline{x}}\right)}{\rmd\underline{\underline{x}}}-\frac{\rmd\ln S\left(\left[\underline{\underline{x}}\right]^{2}\right)}{\rmd\underline{\underline{x}}}\hat{\Lambda}^{N}\left(\underline{\underline{x}}\right).\label{eq:chainrule}
\end{equation}
Using the above identity in (\ref{eq:dQ/dt}) and the definition of
the Majorana Q-function we get:
\begin{eqnarray}
\frac{\rmd Q\left(\underline{\underline{x}}\right)}{\rmd t} & = & -\Omega_{\mu\nu}\left[-x_{\mu\kappa}\frac{\rmd Q}{\rmd x_{\upsilon\kappa}}+x_{\mu\kappa}\frac{\rmd\ln S\left(\left[\underline{\underline{x}}\right]^{2}\right)}{\rmd x_{\upsilon\kappa}}\right]\nonumber \\
 &  & -\Omega_{\mu\nu}\left[\frac{\rmd Q}{\rmd x_{\kappa\mu}}x_{\kappa\upsilon}-\frac{\rmd\ln S\left(\left[\underline{\underline{x}}\right]^{2}\right)}{\rmd x_{\kappa\mu}}x_{\kappa\upsilon}\right].\label{eq:TEv_QFGB}
\end{eqnarray}
Here we have used the following convention for calculation matrix
derivatives $\left[\rmd/\rmd x\right]_{\kappa\upsilon}=\rmd/\rmd x_{\upsilon\kappa}$
\cite{Corney_PD_JPA_2006_GR_fermions}. 

In the limit $S\left(\left[\underline{\underline{x}}\right]^{2}\right)\rightarrow1$,
the above expression reduces to (\ref{eq:TEf}). Next, using the chain
rule we obtain the following expressions: 
\begin{eqnarray}
-x_{\mu\kappa}\frac{\rmd Q}{\rmd x_{\upsilon\kappa}} & = & -\frac{\rmd}{\rmd x_{\upsilon\kappa}}\left(x_{\mu\kappa}Q\right)+\left(\frac{\rmd}{\rmd x_{\upsilon\kappa}}x_{\mu\kappa}\right)Q\nonumber \\
 & = & -\frac{\rmd}{\rmd x_{\upsilon\kappa}}\left(x_{\mu\kappa}Q\right)+\left(2M-1\right)\delta_{\mu\upsilon}Q,\label{eq:CR}
\end{eqnarray}
and

\begin{eqnarray}
\frac{\rmd Q}{\rmd x_{\kappa\mu}}x_{\kappa\upsilon} & = & \frac{\rmd}{\rmd x_{\kappa\mu}}\left(Qx_{\kappa\upsilon}\right)-\left(\frac{\rmd}{\rmd x_{\kappa\mu}}x_{\kappa\upsilon}\right)Q\nonumber \\
 & = & \frac{\rmd}{\rmd x_{\kappa\mu}}\left(Qx_{\kappa\upsilon}\right)-\left(2M-1\right)\delta_{\upsilon\mu}Q.\label{eq:CR-1}
\end{eqnarray}
On substituting (\ref{eq:CR} and \ref{eq:CR-1}) in (\ref{eq:TEf})
we get (\ref{eq:TEf2}). The method of characteristics allow us to
solve the above equation as:

\begin{equation}
\frac{\rmd x_{\upsilon\kappa}}{\rmd t}=\Omega_{\nu\mu}x_{\mu\kappa}-x_{\upsilon\mu}\Omega_{\mu\kappa},
\end{equation}
where we used that $\underline{\underline{\Omega}}$ is an antisymmetric
matrix. In matrix form the above equation is:

\begin{equation}
\frac{\rmd\underline{\underline{x}}}{\rmd t}=\left[\underline{\underline{\Omega}},\underline{\underline{x}}\right].\label{eq:dxdtT-1}
\end{equation}
Here (\ref{eq:dxdtT-1}) corresponds to (\ref{eq:TEX}).

\section{Time evolution for the open quantum system \label{sec:AppTimeEvOpenQS}}

In this section we show the details of the calculations in order to
obtain the time evolution equation of the small quantum dot coupled
to a zero temperature reservoir, for the multimode case. This equation
is given in (\ref{eq:finalEx}). The time evolution equation for the
Q-function written in terms of products of Majorana variables and
Majorana Gaussian operators is given in (\ref{eq:timeevo2}). This
equation is: 
\begin{eqnarray*}
\frac{\rmd Q}{\rmd t} & = & -\frac{i}{2}\frac{1}{{\cal N}}S\Tr\left[\tilde{\Omega}_{\kappa\nu}\Bigl\{\hat{\underline{\gamma}}:\hat{\underline{\gamma}}^{T}\hat{\Lambda}:\Bigr\}_{\nu\kappa}\widehat{\rho}\right]\\
 &  & -\frac{1}{2i}\frac{1}{{\cal N}}S\Tr\left[\Upsilon_{\kappa\nu}\left(:\hat{\underline{\gamma}}\hat{\underline{\gamma}}^{T}\hat{\Lambda}:{}_{\nu\kappa}-\Bigl\{\hat{\underline{\gamma}}:\hat{\underline{\gamma}}^{T}\hat{\Lambda}:\Bigr\}_{\nu\kappa}\right)\widehat{\rho}\right]-\gamma_{ij}\delta_{ij}Q.
\end{eqnarray*}
We now use the Majorana differential identities given in (\ref{eq:MajMixProd})
and (\ref{eq:NormalNormMajId}) respectively. On simplifying terms
and using identities corresponding identities we obtain:
\begin{eqnarray}
\frac{\rmd Q}{\rmd t} & = & \frac{1}{2}\frac{1}{{\cal N}}S\Tr\left[\tilde{\Omega}_{\kappa\nu}\left[\underline{\underline{X}}^{+}\frac{d\hat{\Lambda}\left(\underline{\underline{X}}\right)}{d\underline{\underline{X}}}\underline{\underline{X}}^{-}-\underline{\underline{X}}^{-}\hat{\Lambda}\left(\underline{\underline{X}}\right)\right]_{\nu\kappa}\widehat{\rho}\right]\nonumber \\
 & + & \frac{1}{{\cal N}}S\Tr\left[\Upsilon_{\kappa\nu}\left[\underline{\underline{X}}\frac{d\hat{\Lambda}\left(\underline{\underline{X}}\right)}{d\underline{\underline{X}}}\underline{\underline{X}}^{-}\right]_{\nu\kappa}\widehat{\rho}\right]-\Upsilon_{\kappa\nu}X_{\nu\kappa}Q\label{eq:timeevo3}
\end{eqnarray}
provided $\gamma_{ij}\delta_{ij}=\Upsilon_{\kappa\nu}\mathcal{I}_{\nu\kappa}.$
Due to the same reason as stated in the previous time evolution example,
$\tilde{\Omega}_{\kappa\nu}X_{\nu\kappa}^{-}$ will be zero. Now utilizing
the chain rule (\ref{eq:chainrule}), the definition of the Q-function
and considering the limit $S\left(\left[\underline{\underline{X}}\right]^{2}\right)\rightarrow1,$
we get:

\begin{equation}
\frac{\rmd Q}{\rmd t}=\frac{1}{2}\tilde{\Omega}_{\kappa\nu}\left[\underline{\underline{X}}^{+}\frac{\rmd Q}{\rmd\underline{\underline{X}}}\underline{\underline{X}}^{-}\right]_{\nu\kappa}+\Upsilon_{\kappa\nu}\left[\underline{\underline{X}}\frac{\rmd Q}{\rmd\underline{\underline{X}}}\underline{\underline{X}}^{-}\right]_{\nu\kappa}-\Upsilon_{\kappa\nu}X_{\nu\kappa}Q\label{eq:timeevo4}
\end{equation}

Next, using the following result of the product rule:

\begin{equation}
\fl X_{\nu\ell}\frac{\rmd Q}{\rmd X_{p\ell}}X_{p\kappa}^{-}=\frac{\rmd}{\rmd X_{p\ell}}X_{\nu\ell}X_{p\kappa}^{-}Q-\left(\delta_{\nu p}\delta_{\ell\ell}-\delta_{\nu\ell}\delta_{\ell p}\right)X_{p\kappa}^{-}Q-\left(\delta_{pp}\delta_{\kappa\ell}-\delta_{p\ell}\delta_{\kappa p}\right)X_{\nu\ell}Q,\label{eq:CR2}
\end{equation}
on the time evolution equation we finally obtain:
\begin{eqnarray*}
\frac{\rmd Q}{\rmd t} & = & \frac{1}{2}\tilde{\Omega}_{\kappa\nu}\frac{\rmd}{\rmd X_{p\ell}}\left(X_{\nu\ell}^{+}QX_{p\kappa}^{-}\right)+\Upsilon_{\kappa\nu}\frac{\rmd}{\rmd X_{p\ell}}\left(X_{\nu\ell}X_{p\kappa}^{-}Q\right)-\Upsilon_{\kappa\nu}X_{\nu\kappa}Q\\
 &  & -\Upsilon_{\kappa\nu}\left(\delta_{\nu p}\delta_{\ell\ell}-\delta_{\nu\ell}\delta_{\ell p}\right)X_{p\kappa}^{-}Q-\Upsilon_{\kappa\nu}\left(\delta_{pp}\delta_{\kappa\ell}-\delta_{p\ell}\delta_{\kappa p}\right)X_{\nu\ell}Q.
\end{eqnarray*}
Upon summation by considering all the $2M$ modes, $\sum\delta_{\ell\ell}=2M$,
and on simplifying we finally get:

\begin{eqnarray}
\frac{\rmd Q}{\rmd t} & = & \frac{1}{2}\tilde{\Omega}_{\kappa\nu}\frac{\rmd}{\rmd X_{p\ell}}\left(X_{\nu\ell}^{+}QX_{p\kappa}^{-}\right)+\Upsilon_{\kappa\nu}\frac{\rmd}{\rmd X_{pl}}\left(X_{\nu\ell}QX_{p\kappa}^{-}\right)\nonumber \\
 &  & -\left(4M-1\right)X_{\nu\kappa}\Upsilon_{\kappa\nu}Q+\left(2M-1\right)\Upsilon_{\kappa\nu}\mathcal{I}_{\nu\kappa}.\label{eq:finalEx-1}
\end{eqnarray}
This corresponds to (\ref{eq:finalEx}).

\section*{References}{}

\bibliographystyle{iopart-num}

\end{document}